%% file: text.tex
\documentclass[11pt]{article}
\usepackage{amsmath,epsfig,sint}
\usepackage[american]{babel}
\input commands.tex

\begin{document}
  \input title.tex
  \input section1.tex
  \input section2.tex

  \input section3.tex
  \input section4.tex
  \input section5.tex
  \input section6.tex

  \input section7.tex
  \input acknow.tex
  \input appendixA.tex
  \input appendixB.tex

\input ref.tex
\end{document}

%% file: commands.tex
\def\lvec#1{\setbox0=\hbox{$#1$}
    \setbox1=\hbox{$\scriptstyle\leftarrow$}
    #1\kern-\wd0\smash{
    \raise\ht0\hbox{$\raise1pt\hbox{$\scriptstyle\leftarrow$}$}}
    \kern-\wd1\kern\wd0}
\def\rvec#1{\setbox0=\hbox{$#1$}
    \setbox1=\hbox{$\scriptstyle\rightarrow$}
    #1\kern-\wd0\smash{
    \raise\ht0\hbox{$\raise1pt\hbox{$\scriptstyle\rightarrow$}$}}
    \kern-\wd1\kern\wd0}
\def\lrvec#1{\setbox0=\hbox{$#1$}
    \setbox1=\hbox{$\scriptstyle\leftrightarrow$}
    #1\kern-\wd0\smash{
    \raise\ht0\hbox{$\raise1pt\hbox{$\scriptstyle\leftrightarrow$}$}}
    \kern-\wd1\kern\wd0}

\newcommand{\mcr}{m_{\rm cr}}
\newcommand{\Dslash}{\mbox{$D${\hspace{-1.5ex}/}}\,}
\newcommand{\pslash}{\mbox{$p${\hspace{-1.0ex}/}}\,}

\def\op#1{{\cal O}_{#1}}
\newcommand{\unit}{1\kern-.25em {\rm l}}

\newcommand{\be}{\begin{equation}}
\newcommand{\ee}{\end{equation}}
\newcommand{\bd}{\begin{displaymath}}
\newcommand{\ed}{\end{displaymath}}
\newcommand{\bea}{\begin{eqnarray}}
\newcommand{\eea}{\end{eqnarray}}
\newcommand{\ba}{\begin{array}}
\newcommand{\ea}{\end{array}}

\newcommand{\tr}{{\rm tr}}

\newcommand{\Id}{\mathbf{1}}
\newcommand{\Real}{\relax{\mathsf{\Gamma\kern-.35em R}}}
\newcommand{\Int}{\relax{\mathsf{Z\kern-.40em Z}}}

\newcommand{\half}{{\tfrac{1}{2}}}

\newcommand{\diag}{{\rm diag}}

\newcommand{\Nf}{N_{\rm f}}

\newcommand{\bfp}{{\bf p}}

\newcommand{\bfx}{{\bf x}}
\newcommand{\bfy}{{\bf y}}
\newcommand{\bfz}{{\bf z}}
\newcommand{\bfn}{{\bf n}}

\newcommand{\rmd}{{\rm d}}
\newcommand{\rme}{{\rm e}}
\newcommand{\rmO}{{\rm O}}

\newcommand{\calD}{{\cal D}}

\newcommand{\muq}{\mu_{\rm q}}

\newcommand{\chibar}{\bar{\chi}}
\newcommand{\psibar}{\bar{\psi}}
\newcommand{\rhobar}{\bar{\rho}}
\newcommand{\etabar}{\bar{\eta}}

\newcommand{\zetabar}{\bar{\zeta}}
\newcommand{\zetaprime}{\zeta\kern1pt'}
\newcommand{\zetabarprime}{\zetabar\kern1pt'}
\newcommand{\rhoprime}{\rho'}
\newcommand{\rhobarprime}{\rhobar'}

\newcommand{\msbar}{{\rm \overline{MS\kern-0.05em}\kern0.05em}}

\def\ca{c_{\rm A}}
\def\cv{c_{\rm V}}
\def\csw{c_{\rm sw}}

\def\rmd{{\rm d}}

%% file: title.tex
\begin{titlepage}
\begin{flushright}
   TCDMATH 10--07
\end{flushright}
\vskip 0.5cm
\begin{center}
  {\Large\bf  The chirally rotated Schr\"odinger functional with Wilson fermions and automatic O($a$) improvement\\[1.5ex]}
\end{center}
\vskip 1cm
\begin{center}
{\large Stefan Sint}
\vskip 2.3ex

School of Mathematics\\
Trinity College Dublin\\
Dublin 2, Ireland\\

\vskip 1.5cm

{\bf Abstract}
\vskip 0.7ex
\end{center}
A  modified formulation of the Schr\"odinger functional (SF) is proposed. In the continuum it
is related to the standard SF by a non-singlet chiral field rotation and therefore referred to
as the chirally rotated SF ($\chi$SF). On the lattice with Wilson fermions
the relation is not exact, suggesting some interesting tests of universality. 
The main advantage of the $\chi$SF consists in its compatibility with the mechanism 
of automatic O($a$) improvement. In this paper the basic set-up is introduced and discussed. 
Chirally rotated SF boundary conditions are implemented on the lattice using an orbifold construction. 
The lattice symmetries imply a list of counterterms, which determine how the action and 
the basic fermionic two-point functions are renormalised and O($a$) improved.
As with the standard SF, a logarithmically divergent boundary counterterm leads to a multiplicative 
renormalisation of the fermionic boundary fields. In addition, a finite dimension 3 boundary counterterm  
must be tuned in order to preserve the chirally rotated boundary conditions in the interacting theory. 
Once this is achieved, O($a$) effects originating from the bulk action or from insertions of 
composite operators in the bulk can be avoided by the mechanism of automatic O($a$) improvement. 
The remaining O($a$) effects  arise from the boundaries and can be cancelled by tuning a couple 
of O($a$) boundary counterterms.  The general results are  illustrated in the free theory where the 
Sheikholeslami-Wohlert term is shown to affect correlation functions only at O($a^2$), 
irrespective of its coefficient.
\vspace{1cm}
\begin{center}
  August 2010, revised January 2011
\end{center}
\vfill
\eject

\end{titlepage}

%% file: section1.tex
\section{Introduction}
The Schr\"odinger functional (SF)~\cite{Luscher:1992an-Sint:1995rb} has become a general 
tool to address non-perturbative renormalization problems in lattice 
QCD~\cite{Jansen:1995ck}\footnote{see \cite{Sint:2000vc} for early references 
and \cite{DellaMorte:2005kg-Sommer:2006sj} for a selection of more recent applications 
and further references.}  and is now even used in candidate theories which might 
describe a strongly interacting electroweak 
sector~\cite{Appelquist:2009ty-Yamada:2010wd}.
Renormalization schemes based on the SF are gauge invariant, quark mass independent 
through renormalization in the chiral limit and suitable for evaluation by both Monte Carlo and 
perturbative methods. Moreover, as the finite space-time volume is used
to set the renormalization scale, recursive finite size techniques can be applied to 
bridge large scale differences, thereby avoiding to resolve widely different scales on a single lattice. 
As systematic errors in large scale lattice simulations are becoming smaller, 
the need for  a better controlled continuum limit will likely stir more interest in SF schemes, 
and it is therefore necessary to address its remaining shortcomings.

A first problem consists in the difficulty to implement the SF boundary conditions for
fermion regularisations other than Wilson fermions. 
This has been largely solved, and formulations of the SF
with staggered and Ginsparg-Wilson-type fermions have been given in 
\cite{Heller:1997pn-PerezRubio:2008yd}
and \cite{Taniguchi:2004gf-Sint:2007zz}, respectively.

The main problem with SF schemes is the presence of lattice artefacts which are
linear in the lattice spacing $a$.
Some of these O($a$) effects are caused by the mere presence of the Euclidean time boundaries,
together with local Dirichlet conditions for the fields. Such O($a$) boundary effects will
be present with any regularisation, since they do not arise from the breaking of a continuum symmetry.
In practical applications, these effects are cancelled by adding
a couple of boundary counterterms to the action, with coefficients determined perturbatively 
up to two-loop order~\cite{Bode:1998hd,Bode:1999sm}.  While this is often sufficient in practice, 
a non-perturbative determination of such boundary counterterms would be desirable 
and is indeed conceivable.

With Wilson quarks there is a second category of O($a$) effects which are cancelled by 
the usual O($a$) counterterms  to the Wilson quark action and the composite fields which appear 
in the correlation functions. At first sight it may be surprising that these standard O($a$) 
counterterms are required at all, given that massless Wilson quarks in a finite volume enjoy the property of 
automatic O($a$) improvement~\cite{Frezzotti:2003ni}. However, the argument for automatic O($a$) improvement  
relies on a discrete chiral symmetry which is expected to be recovered in the continuum limit. 
The argument fails as the standard SF boundary conditions break chiral symmetry, 
so that observables cannot be classified as either even or odd under this 
symmetry~\cite{Sint:2005qz,Sint:2007ug}.

In this paper a modified definition of the Schr\"odinger functional for Wilson-type quarks is introduced, 
which is suitable for QCD with an even number of quark flavours, and which 
is compatible with automatic O($a$) improvement. In the continuum limit this modified SF 
is related to the standard SF by a chiral field rotation and is therefore 
referred to as the chirally rotated SF ($\chi$SF). 
A first account of this work has been given a while ago in refs.~\cite{Sint:2005qz,Sint:2007ug}. 
The paper is organised as follows: in Section~2 the argument of automatic O($a$) improvement 
is reviewed and the reason why it fails in the presence of SF boundary conditions.
Possible modifications of the SF boundary conditions which may restore this
argument are discussed next. Concentrating on the solution provided by the chirally rotated SF,
its lattice implementation through an orbifold construction is explained in Section~4. 
The lattice symmetries imply the counterterm structure relevant 
for renormalisation and O($a$) improvement (Section~5). In Section~6,  
the renormalisation constants and boundary O($a$) improvement coefficients 
are determined to tree-level of perturbation theory, 
and automatic bulk O($a$) improvement is demonstrated to this order. 
The paper ends with some concluding remarks and an outlook (Section~7). Finally, Appendix~A
demonstrates the absence of zero modes in the continuum Dirac operator and Appendix B collects
some useful formulae for the free fermion propagator both on and off the lattice.

%% file: section2.tex
\section{Automatic O($a$) improvement and SF boundary conditions}

In order to understand how the SF boundary conditions interfere with O($a$) improvement, it is useful
to recall the argument why massless Wilson fermions are expected to be automatically O($a$) improved in a 
finite volume without boundaries.

\subsection{Symanzik's effective theory and automatic O($a$) improvement}

The Wilson fermion action is given by
\begin{eqnarray}
 S_f &=& a^4\sum_x \bar\psi(x)(D_{\rm W}+m_0)\psi(x), \\
 D_{\rm W} &=& \sum_{\mu=0}^3
  {\textstyle\frac{1}{2}}\left\{(\nabla_{\mu}^{}+\nabla_{\mu}^{\ast})\gamma_\mu
    -a\nabla_{\mu}^{\ast}\nabla_{\mu}^{}\right\} + \csw \frac{i}{4} a \sum_{\mu,\nu=0}^3\sigma_{\mu\nu}\hat{F}_{\mu\nu}(x).
\label{eq:Dw}
\end{eqnarray}
Here, $m_0$ is a bare mass parameter, the covariant lattice derivatives
in the Wilson-Dirac operator are defined as usual, and the last term is the Sheikholeslami-Wohlert 
term with $\hat{F}_{\mu\nu}$ denoting the clover leaf definition 
of the field strength tensor (see ref.~\cite{Luscher:1996sc} for conventions and unexplained notation).
Standard Wilson quarks are obtained for $\csw=0$. We will focus on the massless limit,
which is obtained by tuning the bare mass parameter to its critical value, $m_0=\mcr$.

Cutoff effects can be studied using Symanzik's effective continuum 
theory~\cite{Symanzik:1983dc,Luscher:1996sc}. 
The effective action,
\begin{equation}
 S_{\rm eff}  = S_0+aS_1+ a^2 S_2 +\rmO(a^3),
\end{equation}
starts out with the massless QCD continuum action,
\begin{equation}
 S_0 = -\frac{1}{2g^2}\int{\rm d}^4x\ \tr \left\{F_{\mu\nu}(x)F_{\mu\nu}(x)\right\} 
         + \int{\rm d}^4x\ \psibar(x)\Dslash\psi(x).
\end{equation}
The terms which may appear in $S_1$ must be integrals over 
local composite fields of mass dimension 5.
In the chiral limit there are only two such terms, 
\begin{equation}
   S_1= i  c_1 \int{\rm d}^4x\ \psibar(x) \sigma_{\mu\nu}F_{\mu\nu}(x)\psi(x) + 
           c_2 \int{\rm d}^4x\ \psibar(x) D_\mu D_\mu \psi(x),
\label{eq:S1}
\end{equation}
which can be reduced to a single term by restricting attention to on-shell quantities.
The effective action renders the cutoff dependence explicit, 
up to logarithmic terms (cf.~ref.~\cite{Balog:2009np}).
The leading cutoff effects in a renormalised connected correlation 
function of a (multi-local) observable $O$ are then described by the corresponding  
continuum correlation functions with insertions of either $S_1$ or the O($a$) counterterm 
$\delta O$ for the observable $O$,
\begin{eqnarray}
  \langle O\rangle_{\rm c} &=& \langle O\rangle^{\rm cont}_{\rm c}
  -a \langle S_1 O\rangle^{\rm cont}_{\rm c}
  +a \langle \delta O\rangle^{\rm cont}_{\rm c}+  \rmO(a^2).
\label{eq:contlimit}
\end{eqnarray}
Here, $\langle \cdot \rangle^{\rm cont}_{\rm c}$ refers to Euclidean expectation values
taken with respect to the continuum action $S_0$, only taking into account connected diagrams. Obviously, 
the continuum action, $S_0$, is chirally symmetric. In particular, the discrete
$\gamma_5$-transformation
\begin{equation}
   \psi\rightarrow\gamma_5\psi,\qquad 
   \psibar\rightarrow -\psibar\gamma_5,
\label{eq:gam5cont}
\end{equation}
leaves $S_0$ invariant, while it changes the sign of $S_1$.
Since the $\gamma_5$-transformation squares to the identity, one may decompose
any composite field into parts which are either even or odd under this transformation.
Note that, despite the loss of chiral invariance on the lattice, such a decomposition 
can be made unambiguously for any lattice field. An important observation then is
that the O($a$) counterterm, $\delta O_{\rm even}$ for any $\gamma_5$-even 
field $O_{\rm even}$ must be $\gamma_5$-odd  and 
vice versa~\cite{Frezzotti:2003ni}. 
Provided that the functional measure is invariant under the change of variables
to the $\gamma_5$-transformed fields, one is then led to conclude that
\bea
     \langle S_1 O_{\rm even} \rangle^{\rm cont}_{\rm c} 
&=& -\langle S_1 O_{\rm even} \rangle^{\rm cont}_{\rm c}= 0,\\
     \langle \delta O_{\rm even} \rangle^{\rm cont}_{\rm c}  &=& 
     -\langle \delta O_{\rm even} \rangle^{\rm cont}_{\rm c}=0,
\eea
i.e.~Eq.~(\ref{eq:contlimit}) for $\gamma_5$-even observables reads
\begin{equation}
   \langle O_{\rm even}\rangle_{\rm c} = \langle O_{\rm even}\rangle^{\rm cont}_{\rm c} +\rmO(a^2).
\end{equation}
This does not mean that O($a$) effects are absent, rather these are present in correlation functions 
of $\gamma_5$-odd observables $O_{\rm odd}$,
for which the leading term in (\ref{eq:contlimit}) vanishes, so that
\begin{equation}
  \langle O_{\rm odd}\rangle_{\rm c}= -a \langle S_1 O_{\rm odd}\rangle^{\rm cont}_{\rm c}
  + a \langle \delta O_{\rm odd}\rangle^{\rm cont}_{\rm c} + \rmO(a^3).
\end{equation}
Note that the corrections here are of order $a^3$, rather than $a^2$. In fact, the mechanism
of automatic O($a$) improvement more generally implies that lattice effects in $\gamma_5$-even 
($\gamma_5$-odd)  correlation functions only come with even (odd) powers of the lattice spacing 
$a$~\cite{Frezzotti:2003ni,Aoki:2006nv}.

It should be emphasised that the inherent O($a$) ambiguity in the definition of
the chiral limit with Wilson quarks does not alter this picture. To understand this, suppose
that $m_0$ has been tuned critical e.g. by requiring the divergence of the axial current to vanish
for some particular matrix element. By making a different choice, a non-zero quark mass $m'= a\Lambda^2$ 
is generated relative to the previous definition, with $\Lambda$ denoting some mass scale. 
Its effect can be accounted for by an insertion of a fermionic mass term. Since this
term  is $\gamma_5$-odd, its insertion into an even correlation function vanishes,
\begin{equation}
  \int{\rm d}^4 x\,\left\langle O_{\rm even} \psibar(x)\psi(x)\right\rangle^{\rm cont}_{\rm c} = 0.
\end{equation}
In order to obtain a finite contribution to $\gamma_5$-even observables
at least a double insertion of this term is required, making it an $a^2$-effect, as anticipated.

Note that the above arguments rely on the assumption that the functional 
measure is invariant under the change of variables to the $\gamma_5$-transformed fields. 
To check this assumption it is useful to think of the effective Symanzik theory 
as being formulated on a very fine lattice with a spacing $b$ very much smaller than $a$ 
such that the theory is essentially continuum-like.
Nevertheless, the functional integral in a finite volume is still finite-dimensional 
and mathematically well-defined. Furthermore, with Ginsparg-Wilson quarks,
the fermionic part of the continuum like action
\begin{equation}
   S_{0,f}= b^4\sum_x \psibar(x) D_N \psi(x),
\label{eq:GWaction}
\end{equation}
has an exact chiral symmetry~\cite{Luscher:1998pqa}. Here, $D_N$ is the Neuberger operator~\cite{Neuberger:1997fp}, 
which satisfies the Ginsparg-Wilson relation~\cite{Ginsparg:1981bj},
\begin{equation}
   \gamma_5 D_N + D_N \gamma_5=bD_N\gamma_5D_N.
\label{eq:GWrelation}
\end{equation}
Combined with the $\gamma_5$-conjugation property, $D_N^\dagger=\gamma_5 D_N \gamma_5$,
this gives rise to the definition of $\hat\gamma_5$,
\begin{equation}
   \hat\gamma_5 = \gamma_5V,\qquad V=1-bD_N,
\end{equation}
which is hermitian and squares to the identity. However, in contrast to the ordinary $\gamma_5$, 
it depends implicitly on the gauge field through the unitary matrix $V$. 
Replacing the $\gamma_5$-transformation (\ref{eq:gam5cont}) with its version on the fine lattice,
\begin{equation}
    \psi \rightarrow \hat\gamma_5\psi,\qquad \psibar\rightarrow -\psibar\gamma_5,
\label{eq:GW_gam5transf}
\end{equation}
one readily verifies the invariance of the action (\ref{eq:GWaction}), which follows from   
the Ginsparg-Wilson relation (\ref{eq:GWrelation}). 
Changing variables in the functional integral to the $\gamma_5$-transformed fields, 
one finds that the Jacobian $J$ is given by  
\begin{equation}
 J = \det(\gamma_5\hat\gamma_5) = \det(V) = \det(\gamma_5 V \gamma_5) = \det(V^\dagger) = \pm 1.
 \label{eq:jacobian}
\end{equation}
For $\Nf$ quark flavours, each $\gamma_5$-transformed flavour thus contributes a factor $J$,
so that the total Jacobian becomes $J^{\Nf}$. Hence, for even $\Nf$, the $\gamma_5$-transformation 
(\ref{eq:GW_gam5transf}) is an exact symmetry of the path integral 
measure\footnote{For a single quark flavour, a gauge field dependent sign in the Jacobian remains, 
thereby spoiling the argument for automatic O($a$) improvement. 
While some additional argument would be needed for $\Nf=1$, 
the situation is slightly better for odd flavour numbers $\Nf > 1$, 
as the argument can then be applied to any pair of fermion flavours.}.
In conclusion, one expects that lattice QCD with an even number of Wilson type 
quarks is indeed automatically O($a$) improved, 
at least in a finite volume with periodic boundary conditions.

However, the situation changes in the presence of SF boundary conditions, which, 
for the fermionic fields, read~\cite{Sint:1993un} ,
\begin{xalignat}{2}
     P_+\psi(x)\vert_{x_0=0} &= 0,    & P_-\psi(x)\vert_{x_0=T} &=0,
   \nonumber\\
   \psibar(x)P_-\vert_{x_0=0} &= 0,   
    &\psibar(x) P_+\vert_{x_0=T} &= 0, \label{eq:SFbcs}
\end{xalignat}
with $P_\pm=\frac12(1\pm\gamma_0)$.
The $\gamma_5$-transformation is then no longer a symmetry of the effective continuum theory, 
as the $\gamma_5$-transformed fields satisfy the SF boundary conditions with the complementary projectors, e.g.
\begin{equation}
    \psi'=\gamma_5\psi \quad \Rightarrow \quad  P_-\psi'\vert_{x_0=0} = \gamma_5 P_+\psi\vert_{x_0=0} =0.
\end{equation}
This may be taken as a property of the fermion measure, since the function space integrated over is
not the same before and after the change of variables. Hence, the transformed correlation functions
cannot be proportional to the original ones, and the basic argument needed for automatic O($a$)
improvement cannot be made.

\subsection{Rendering automatic O($a$) improvement compatible with the SF}

The question then arises whether there are ways to save automatic
O($a$) improvement in the presence of SF boundary conditions. 
One needs to either find an alternative to the $\gamma_5$-transformation or modify the SF boundary conditions such
that the transformation remains a symmetry of the massless continuum action,
while the (homogeneous) SF boundary conditions are left invariant.
There are various ways to achieve this:
\begin{itemize}
\item
Let $\psi$ be a flavour doublet, and augment the $\gamma_5$-transformation 
by a flavour permutation, i.e.
\begin{equation}
    \psi \rightarrow \tau^1\gamma_5\psi,\qquad \psibar\rightarrow-\psibar\gamma_5\tau^1,
 \label{eq:gam5tau1}
\end{equation}
where $\tau^1$ is a Pauli matrix. This flavour permutation does not affect the bulk action 
and O($a$) improvement for $\gamma_5\tau^1$-even observables in the bulk can be shown as before.
Then, giving the boundary conditions a flavour structure through the replacement
\begin{equation}
   P_\pm=\tfrac12(1\pm\gamma_0) \quad\longrightarrow \quad {\tilde P}_\pm=\tfrac12(1\pm\gamma_0\tau^3),
\end{equation}
one easily sees that, due to 
\begin{equation}
 [\gamma_5\tau^1,{\tilde P}_\pm]=0, 
\end{equation}
the boundary conditions are preserved by the $\gamma_5\tau^1$-transformation. 

\item Another possibility consists in combining the $\gamma_5\tau^1$-transformation (\ref{eq:gam5tau1})
with Dirichlet boundary conditions using the projectors
\begin{equation}
   \tilde{Q}_\pm = \tfrac12(1\pm i\gamma_0\gamma_5\tau^3) \quad \Rightarrow \quad
   [\gamma_5\tau^1,\tilde{Q}_\pm]=0. 
 \label{eq:tildeqpm}
\end{equation}
It is this option which will be investigated in much more detail below.
\end{itemize}
While it is relatively easy to find alternative SF boundary conditions and/or alternatives
to the $\gamma_5$-transformation, the difficult part is to show that this leads 
to a sensible alternative definition of the Schr\"odinger functional in the continuum limit. 
Besides renormalizability one would like to maintain the properties of 
the standard SF such as the absence of fermionic zero modes, which is crucial for 
practical applications. Another difficulty lies in the actual implementation 
of these boundary conditions on the lattice.
All these questions will be addressed in the following for the proposed solution 
involving the projectors in Eq.~(\ref{eq:tildeqpm}).

%% file: section3.tex
\section{Continuum considerations}

The boundary conditions involving the projectors $\tilde{Q}_\pm$   
can be thought of as a chirally rotated version of the standard SF boundary conditions.
As the non-singlet chiral rotation is part of the continuum symmetries of 
massless QCD, the alternative formulation of the SF to be introduced here 
will in fact be closely related to the standard SF. It is useful to have a closer look
at this relationship before passing to the lattice formulation.

\subsection{Chirally rotated SF boundary conditions}

Let $\psi'$ and $\psibar'$ be isospin doublets  
of quark and anti-quark fields satisfying the homogeneous 
SF boundary conditions, Eq.(\ref{eq:SFbcs}).
Performing a non-singlet chiral field rotation,
\begin{equation}
    \psi'= R(\alpha)\psi,\qquad \psibar'=\psibar R(\alpha), \qquad 
     R(\alpha)=\exp(i\alpha \gamma_5\tau^3/2),
 \label{eq:chiral_rot}
\end{equation}
the rotated fields satisfy the chirally rotated boundary conditions,
\begin{xalignat}{2}
     P_+(\alpha)\psi(x)\vert_{x_0=0} &=0,    
    &P_-(\alpha)\psi(x)\vert_{x_0=T} &=0,\nonumber\\
   \psibar(x)\gamma_0P_-(\alpha)\vert_{x_0=0} &= 0,   
    &\psibar(x)\gamma_0P_+(\alpha)\vert_{x_0=T} &= 0, 
\label{eq:rotbc}
\end{xalignat}
with the projectors
\begin{equation}
  P_\pm(\alpha)=\frac12\left[1\pm\gamma_0\exp(i\alpha\gamma_5\tau^3)\right].
\end{equation}
For $\alpha=0$ the standard projectors $P_\pm=P_\pm(0)$ are recovered,
while for $\alpha=\pi/2$ one finds,
\begin{equation}
   P_\pm(\pi/2)\equiv \tilde{Q}_\pm=\frac12(1\pm i\gamma_0\gamma_5\tau^3).
\end{equation}
Since $\gamma_0 \tilde{Q}_\pm=\tilde{Q}_\mp\gamma_0$, the boundary
conditions (\ref{eq:rotbc}) in this special case read
\begin{xalignat}{2}
     \tilde{Q}_+\psi(x)\vert_{x_0=0} &=0,    
    &\tilde{Q}_-\psi(x)\vert_{x_0=T} &=0,\nonumber\\
   \psibar(x)\tilde{Q}_+\vert_{x_0=0} &= 0,   
    &\psibar(x)\tilde{Q}_-\vert_{x_0=T} &= 0.
\label{eq:bc_rotate}
\end{xalignat}
We thus see that boundary conditions involving the projectors $\tilde{Q}_\pm$ 
appear naturally when chirally rotating quark and anti-quark fields 
which satisfy the standard SF boundary conditions.

\subsection{Properties of the continuum Dirac operator}

In the continuum the rotation angle $\alpha$ parameterizes a 
family of linear spaces of functions
which satisfy homogeneous Dirichlet boundary conditions
in Euclidean time. In the spatial directions we assume $L$-periodicity and we
allow for a constant U(1) gauge background field $i\theta_\mu/L$.
Formally this field may be included in the Dirac operator 
by promoting the SU($N$) colour gauge field to a U($N$) gauge field as follows:
\be
   A_\mu(x)= \sum_{a=0}^{N^2-1} A_\mu^a(x) T^a,\qquad  A_\mu^{0}(x) T^0 =  i\frac{\theta_\mu}{L} \Id.
\ee
Following~\cite{Sint:1995ch,Luscher:1996sc}, we set $\theta_1=\theta_2=\theta_3=\theta$ and $\theta_0=0$.  
With respect to the natural inner product,
\begin{equation}
    (\varphi,\chi) = \int_0^T\rmd x_0\int\rmd^3\bfx\, \varphi(x)^\dagger \chi(x),
\end{equation}
the Dirac operator for a quark doublet and generic angle $\alpha$ is $\gamma_5$-hermitian up to a flavour 
exchange, i.e.
\begin{equation}
   \left(\gamma_5\tau^1\left[\Dslash+m+i\muq\gamma_5\tau^3\right]\right)^\dagger = 
         \gamma_5\tau^1\left[\Dslash+m+i\muq\gamma_5\tau^3\right].
\end{equation}
For later reference both a standard and a twisted mass term have been included.
Hence, a well-defined eigenvalue problem is obtained for any value\footnote{As an aside 
note that for $\alpha=\pi/2$ the massless operator $i\Dslash$ is hermitian for a single fermion
with boundary conditions corresponding to either the up or the down flavour in Eq.~(\ref{eq:bc_rotate}).} 
of $\alpha$.
Denoting the  eigenfunctions in the standard SF by $v_n$, the eigenvalue equation reads
\begin{equation}
   \gamma_5\tau^1\left[\Dslash+m+i\muq\gamma_5\tau^3\right] v_n = \lambda_n v_n.
\end{equation}
Performing the chiral rotation as in Eq.(\ref{eq:chiral_rot}), 
\begin{equation}
    v_n = R(\alpha)w_n
\end{equation}
and using 
\begin{eqnarray}
  R(\alpha)\left[\Dslash+m+i\muq\gamma_5\tau^3\right] R(\alpha) &=& 
   \Dslash+\tilde{m}+i\tilde{\muq}\gamma_5\tau^3,
\end{eqnarray}
one finds,
\begin{equation}
 \gamma_5\tau^1 \left[\Dslash+\tilde{m}+i\tilde{\muq}\gamma_5\tau^3\right] w_n = \lambda_n w_n.
\end{equation}
Hence, the eigenfunctions are related by a chiral rotation, and the eigenvalues are exactly the same
as in the standard SF, provided the mass parameters either vanish or are transformed covariantly,
according to
\begin{equation}
 \tilde{m} =
 m\cos\alpha -\mu_q\sin\alpha,
 \qquad
 \tilde{\mu_q} =
 m\sin\alpha +\mu_q\cos\alpha.
\label{eq:cov_mass}
\end{equation}
Moreover, the absence of zero modes in the massless theory can be established following 
L\"uscher~\cite{Luscher:2006df}, and some details have been deferred to Appendix~A.
Given the absence of zero modes, the free continuum quark propagator can then be 
calculated using standard techniques. Some results are collected in Appendix~B.

\subsection{Relating correlation functions at different values of $\alpha$}

The chiral rotation (\ref{eq:chiral_rot}) is not anomalous, so that
one may perform a corresponding change of variables in the functional integral. 
One obtains relations between correlation functions
defined with the standard and the chirally rotated SF boundary conditions.
This is similar to and generalises the relations between correlation functions in 
twisted mass and standard QCD~\cite{Frezzotti:2000nk,Sint:2007ug}. 
In particular, one expects that these formal continuum relations will hold up to cutoff effects, 
for appropriately renormalised correlation functions on the lattice.

It is instructive to include both a standard
and a twisted mass term in the fermionic action at $\alpha=0$, 
\begin{equation}
 S_f[m,\muq] = \int_0^T{\rm d}x_0 \int_0^L{\rm d}^3{\bf x}\,\,
  \psibar'(x)(\Dslash+m+i\muq\gamma_5\tau^3)\psi'(x),
\end{equation}
even though the focus will later be on massless QCD.
Performing the change of variables, Eq.~(\ref{eq:chiral_rot}), in the Euclidean functional integral
one obtains the generic identities,
\begin{eqnarray}
 \left\langle O[\psi,\psibar]\right\rangle_{(m,\muq,P_{\pm})}=
 \left\langle O[R(\alpha)\psi,\psibar R(\alpha)]\right\rangle_{
        (\tilde{m},\tilde{\mu}_q,P_{\pm}(\alpha))},
\label{eq:equivSF}
\end{eqnarray}
where the integration variables are assumed to be $\psi$ and $\psibar$ 
on both sides of the equation.  In addition to the boundary conditions for $\psi$ at $x_0=0,T$,
the subscript to the correlation functions indicates the mass parameters in the action and
the mass parameters are related as in Eq.~(\ref{eq:cov_mass}).
The relation~(\ref{eq:equivSF}) establishes a mapping of correlation functions
defined at different values of $\alpha$.  An equivalent form of this mapping is given by
\begin{equation}
 \left\langle O[\psi,\psibar]\right\rangle_{(\tilde{m},\tilde{\mu}_{\rm q},\tilde{Q}_{\pm})}=
 \left\langle O[R(-\alpha)\psi,\psibar R(-\alpha)]
      \right\rangle_{(m,\muq,P_{\pm}(\frac{\pi}{2}-\alpha))}.
\label{eq:equivSFp}
\end{equation}
Setting the twisted mass term to zero in the
chirally rotated  SF, and choosing $\alpha=\pi/2$, this simplifies to
\begin{equation}
 \langle O[\psi,\psibar]\rangle_{(m,0,\tilde{Q}_{\pm})}=
 \langle O[R(-\pi/2)\psi,\psibar R(-\pi/2)]\rangle_{(0,-m,P_\pm)},
\end{equation}
which shows that the standard mass parameter in the chirally rotated SF
is equivalent to a (negative) twisted mass parameter in the standard SF.

The simplest fermionic correlation function is the quark propagator. 
Using the explicit expressions of Appendix~B,
one may easily verify the identity (\ref{eq:chiral_rot}) in the free theory, viz.
\begin{equation}
   \left\langle \psi(x)\psibar(y) \right\rangle_{(m,\muq,P_\pm)}^{(g=0)} =
   R(\alpha) \left\langle \psi(x)\psibar(y) \right\rangle^{(g=0)}_{(\tilde{m},\tilde{\mu}_{\rm q},P_\pm(\alpha))} R(\alpha).
\end{equation}
Gauge invariant correlation functions corresponding 
to the ones defined in the standard SF~\cite{Luscher:1996sc,Luscher:1996vw} 
can be defined as well. For this we need the fermionic boundary fields, 
$\zeta,\zetaprime$ and $\zetabar,\zetabarprime$, which are naturally included in the mapping~(\ref{eq:equivSF})  
by identifying them with the non-Dirichlet components near the time boundaries.
More precisely we set
\begin{xalignat}{2}
     \zeta(\bfx) &=  P_-(\alpha)\psi(0,\bfx),    
    &\zetaprime(\bfx) &= P_+(\alpha)\psi(T,\bfx),\nonumber\\
     \zetabar(\bfx) &= \psibar(0,\bfx)P_+(-\alpha), 
    &\zetabarprime(\bfx)&= \psibar(T,\bfx) P_-(-\alpha),  
\label{eq:zeta_alpha}
\end{xalignat}
thus leaving the $\alpha$-dependence implicit.
This should not lead to any confusion, as the fermionic boundary fields mostly appear 
as insertions into correlation functions with the subscripts indicating the projectors 
used for the boundary conditions.

\subsection{Symmetries and conventions}

The standard SF boundary conditions break all chiral symmetries, as 
is best illustrated by the Ward identities~\cite{Luscher:2006df}.
With the global axial and vector non-singlet variations,
\begin{xalignat}{2}
     \delta_{\rm A}^a\psi &= \gamma_5\frac{\tau^a}{2}\psi,    
    &\delta_{\rm V}^a\psi &= \frac{\tau^a}{2}\psi, \nonumber\\
     \delta_{\rm A}^a\psibar &= \psibar \gamma_5\frac{\tau^a}{2}, 
    &\delta_{\rm V}^a\psibar &= -\psibar \frac{\tau^a}{2}, 
\end{xalignat}
the Ward identities take the form,
\begin{equation}
  \left\langle \delta_{\rm X}^a O\right\rangle_{(P_\pm)} =  
  \left\langle \delta_{\rm X}^a S O\right\rangle_{(P_\pm)},\qquad {\rm X}={\rm A,V},
\label{eq:wardgeneric}
\end{equation}
where $O$ denotes an arbitrary composite field, and the subscript to the expectation values
indicates the boundary conditions for $\psi$ at $x_0=0,T$.
The variation of the action is given by
\begin{equation}
   \delta_{\rm X} S = -\int\rmd^4x\, \partial_\mu X_\mu^a(x),
\end{equation}
with the Noether currents
\begin{equation}
  A_\mu^a(x) = \psibar(x)\gamma_\mu\gamma_5\frac{\tau^a}{2}\psi(x),\qquad
  V_\mu^a(x) = \psibar(x)\gamma_\mu\frac{\tau^a}{2}\psi(x).
\end{equation}
The boundary conditions in the spatial directions being periodic, the integral over the total divergence
yields only surface terms in the time direction. The r.h.s.~of Eq.~({\ref{eq:wardgeneric})  becomes
\begin{equation}
  \left\langle \delta_{\rm X}^a S O\right\rangle_{(P_\pm)} = 
  \int\rmd^3{\bfx} \left\langle \left[X_0^a(0,\bfx)-X_0^a(T,\bfx)\right] O\right\rangle_{(P_\pm)}.
\end{equation}
Using the standard SF boundary conditions, Eqs.~(\ref{eq:SFbcs}) and the identity $\gamma_0 P_\pm = \pm P_\pm$, 
one obtains
\begin{eqnarray}
   A_0^a(0,x) &=&  \psibar(0,\bfx)P_+\gamma_5 \frac{\tau^a}{2}P_-\psi(0,\bfx) 
                 = \zetabar(\bfx)\gamma_5 \frac{\tau^a}{2}\zeta(\bfx),\\
   A_0^a(T,x) &=&  \psibar(T,\bfx)P_-\gamma_5 \frac{\tau^a}{2}P_+\psi(0,\bfx) 
                 = \zetabarprime(\bfx)\gamma_5 \frac{\tau^a}{2}\zetaprime(\bfx),
\end{eqnarray}
whereas $V_0^a(0,\bfx)$ and $V_0^a(T,\bfx)$ vanish. One thus arrives at the Ward identities,
\begin{eqnarray}
 \langle \delta_{\rm A}^a O\rangle_{(P_\pm)} &=& 
    \half\int \rmd^3z\left\langle \left[\zetabar(\bfz)\gamma_5\tau^a\zeta(\bfz) +
          \zetabarprime(\bfz)\gamma_5\tau^a\zetaprime(\bfz)\right] O\right\rangle_{(P_\pm)},\\
    \langle  \delta_{\rm V}^a O\rangle_{(P_\pm)} &=& 0,
\end{eqnarray}
which show that the standard SU(2) vector symmetry is conserved, 
while chiral symmetry is broken by terms which are localized at the time boundaries, with coefficients
of O($1$). 

Proceeding analogously we may derive the Ward identities in the presence of
chirally rotated SF boundary conditions,
\begin{eqnarray}
 \langle \delta_{\rm A}^a O\rangle_{(P_\pm(\alpha))}\! &=& \!
   \half\! \int \rmd^3z\left\langle \left[\zetabar(\bfz)\gamma_0\gamma_5\tau^a\zeta(\bfz) -
          \zetabarprime(\bfz)\gamma_0\gamma_5\tau^a\zetaprime(\bfz)\right] O\right\rangle_{(P_\pm(\alpha))},
\label{eq:AWI}\\
  \langle \delta_{\rm V}^a O\rangle_{(P_\pm(\alpha))} \!&=& \!
   \half\! \int \rmd^3z\left\langle \left[\zetabar(\bfz)\gamma_0\tau^a\zeta(\bfz) -
          \zetabarprime(\bfz)\gamma_0\tau^a\zetaprime(\bfz)\right] O\right\rangle_{(P_\pm(\alpha))}.
\label{eq:VWI}
\end{eqnarray}
Using the projector structure implicit in the fermionic boundary fields, Eq.~(\ref{eq:zeta_alpha}), 
the r.h.s.~of the vector Ward identity (\ref{eq:VWI}) with index $a=3$ is seen to vanish. 
Moreover, there are two further linear combinations of Eqs.~(\ref{eq:AWI},\ref{eq:VWI}) for which 
the r.h.s.~vanishes, so that three generators correspond to conserved symmetries.
This is not surprising, as the  chiral rotation merely transforms the vector symmetry 
of the standard SF to an unusual form, a situation which is familiar 
from twisted mass QCD~\cite{Frezzotti:2000nk,Sint:2007ug}.
To illustrate this further, consider the case of most interest, $\alpha=\pi/2$, where the 
Ward identities take the form
\begin{eqnarray}
 \langle \delta_{\rm A}^{a} O\rangle_{(\tilde{Q}_\pm)} &=& \tfrac{i}{2}\delta^{3a}
           \int \rmd^3z\left\langle \left[\zetabar(\bfz)\zeta(\bfz) +
          \zetabarprime(\bfz)\zetaprime(\bfz)\right] O\right\rangle_{(\tilde{Q}_\pm)},
\label{eq:ward1}\\
  \langle \delta_{\rm V}^{a} O\rangle_{(\tilde{Q}_\pm)} &=& \tfrac{i}{2}\varepsilon^{3ab}
    \int \rmd^3z\left\langle \left[\zetabar(\bfz)\gamma_5\tau^{b}\zeta(\bfz) -
          \zetabarprime(\bfz)\gamma_5\tau^b\zetaprime(\bfz)\right] O\right\rangle_{(\tilde{Q}_\pm)}.
\label{eq:ward2}
\end{eqnarray}
Apparently a mixture of vector and axial symmetries are broken by 
boundary terms. However, in massless QCD this is really a convention dependent statement.
A preferred or ``physical" basis is only available in the presence of a mass term: 
those symmetry transformations which leave the mass term invariant are identified as 
vector symmetries whereas the remaining chiral flavour transformations are classified as axial symmetries. 
In massless QCD with SF type boundary conditions we will choose a similar convention, and define
the ``physical" basis as the one with standard SF boundary conditions. This implies that 
the broken symmetries are in all cases the axial symmetries, whereas the vector or isospin symmetries
are conserved. The situation is then completely analogous to twisted mass QCD, in particular, the ``dictionary"
translating composite fields between both bases can be taken over from twisted mass QCD.

%% file: section4.tex
\section{$\chi$SF boundary conditions on the lattice}

Implementing boundary conditions on the lattice is not straightforward, since
boundary conditions cannot really be imposed. Rather, they arise dynamically from the
structure of the lattice action near the boundaries. An attractive, though not
universally applicable technique to obtain a lattice theory with the
desired boundary conditions is based on an orbifold construction. 
In the context of lattice QCD orbifold techniques 
have previously been  applied by Taniguchi~\cite{Taniguchi:2004gf}  
in order to implement SF boundary conditions for Ginsparg-Wilson quarks. 
The r\^ole of the orbifold construction is to define the action of the lattice operator 
near the time boundaries, such that the desired boundary conditions are obtained for the free theory. 
In a second step one needs to analyse the symmetries and list the allowed boundary counterterms 
which will be generated by the interactions. From the list of possible counterterms 
one may conclude whether the desired boundary conditions will be obeyed
in the renormalised theory, or whether this requires the fine tuning 
of some counterterm coefficients (cf.~Sect.~5).

\subsection{Orbifold construction}

The starting point is a single Wilson quark flavour $\chi$ with lattice action
\begin{equation}
  S_f[\chi,\chibar,U] = a^4\!\!\!\sum_{-T< x_0\leq T}
  \sum_{\bf x}\chibar(x)\left(D_W+m_0\right)\chi(x).
  \label{eq:orb-start}
\end{equation}
The fermionic fields are taken to be $2T$-anti-periodic in the Euclidean time
direction,
\begin{equation}
   \chi(x_0+2T,{\bf x})=-\chi(x),\qquad
   \chibar(x_0+2T,{\bf x})=-\chibar(x).
\end{equation}
The orbifold reflection ${\cal T}$ about the Euclidean time $x_0=0$ is defined by,
\begin{equation}
   {\cal T}: \chi(x) \rightarrow ({\cal T}\chi)(x)=
        i\gamma_0\gamma_5\chi(-x_0,{\bf x}),
\qquad
      \chibar(x) \rightarrow (\chibar {\cal T})(x) =
      \chibar(-x_0,{\bf x})i\gamma_0\gamma_5.
\label{eq:orbifold}
\end{equation}
Following ref.~\cite{Taniguchi:2004gf} the gauge field is treated as
an external field. It is first extended to the doubled time interval
$[-T,T]$, through
\begin{equation}
   U_k(-x_0,{\bf x})=U_k(x_0,{\bf x}),\qquad 
   U_0(-x_0-a,{\bf x})^\dagger=U_0(x),
\label{eq:reflectedgaugefield}
\end{equation}
and then $2T$-periodically continued to all Euclidean times.
Since ${\cal T}$ is an involution, one may define
associated projectors and project on ${\cal T}$-even and -odd fields as follows:
\begin{equation}
   \chi_\pm = \frac12(1\pm {\cal T}) \chi,\qquad
   \chibar_\pm=\chibar\frac12(1\pm {\cal T}).
\end{equation}
Even and odd fields then satisfy homogeneous Dirichlet conditions at $x_0=0$,
\begin{equation}
 (1\mp i\gamma_0\gamma_5)\chi_\pm(0,{\bf x})=0,
 \qquad \chibar_\pm(0,{\bf x})(1\mp i\gamma_0\gamma_5) = 0.
\end{equation}
Furthermore, due to the $2T$-anti-periodicity in time,
Dirichlet conditions with the complementary projectors are obtained
at $x_0=T$,
\begin{equation}
  (1\pm i\gamma_0\gamma_5)\chi_\pm(T,{\bf x})=0,
  \qquad \chibar_\pm(T,{\bf x})(1\pm i\gamma_0\gamma_5) = 0.
\end{equation}
${\cal T}$-even and odd anti-quark fields are defined analogously.
For the orbifold projection to be consistent with the lattice theory, it is necessary that
the reflection ${\cal T}$ commutes (or anti-commutes) with the Wilson-Dirac operator.
The lattice action can then be decomposed according to 
\begin{equation}
  S_f[\chi,\chibar,U] =
  S_f[\chi_++\chi_-,\chibar_++\chibar_-,U]
  = S_f[\chi_+,\chibar_+,U] + S_f[\chi_-,\chibar_-,U].
\end{equation}
To check whether this is indeed the case for the reflection~(\ref{eq:orbifold}), the crucial relations to note are 
\begin{equation}
 {\cal T} \nabla_0^\ast = -\nabla_0^{} {\cal T},\qquad 
 {\cal T} \nabla_0^{}   = -\nabla_0^\ast {\cal T},
\end{equation}
for the forward and backward covariant time derivatives.
On the other hand, all spatial derivatives simply commute with ${\cal T}$. 
Taking into account the $\gamma$-matrix structure in the
Wilson-Dirac operator this then leads to $[{\cal T},D_W]=0$, as required.

As a consequence, one may consistently restrict attention to either
the ${\cal T}$-even or the ${\cal T}$-odd fields. For instance, using only the ${\cal T}$-odd fields, 
the fermionic functional integral is taken to be 
\be
   \int D[\chi_-,\chibar_-]\exp(- S_f[\chi_-,\chibar_-,U]).
 \label{eq:oddint}
\ee
Both ${\cal T}$-even and ${\cal T}$-odd fields are determined for negative times
$x_0\in [-T,0]$, once specified for $x_0\in [0,T]$. 
Hence, the field values at the negative times are no independent degrees of freedom,
and the integration measure over the odd fields in Eq.~(\ref{eq:oddint}) can be written as
a measure over the fields at Euclidean times $x_0\in [0,T]$,
\begin{equation}
   D[\chi_-,\chibar_-] =  \prod_{\bfx} \rmd\varphi(\bfx) \rmd\bar\varphi(\bfx) 
                     \prod_{0 < x_0 < T} \rmd\chi_-(x) \rmd\chibar_-(x),
\end{equation}
where the non-Dirichlet components have been combined in the spinors
$\varphi(\bfx)$ and $\bar\varphi(\bfx)$, viz.
\begin{eqnarray}
      \varphi(\bfx) &=& Q_-\chi_-(0,\bfx)+Q_+\chi_-(T,\bfx), \\
      \bar\varphi(\bfx) &=& \chibar_-(0,\bfx)Q_- +\chibar_-(T,\bfx)Q_+,
\end{eqnarray}
with $Q_\pm=\half(1\pm i\gamma_0\gamma_5)$. Also the action can be reduced to 
the interval $[0,T]$: taking into account Eq.~(\ref{eq:reflectedgaugefield}),
the action for the fields at negative times $x_0\in [-T,0]$ turns out to be 
the same as for the fields at positive times. 
Eventually, the only trace left by the orbifold reflection 
is the structure of the Wilson-Dirac operator near the time boundaries.

The case of a flavour doublet with boundary conditions given 
in terms of the projectors $\tilde{Q}_\pm$, is naturally obtained by interpreting 
the even and odd fields as the components of a flavour doublet.
More precisely, one sets
\begin{equation}
  \psi=\sqrt{2}
  {\begin{pmatrix}\chi_-\\ \chi_+ \end{pmatrix}},\qquad
  \psibar= \sqrt{2}
  {\begin{pmatrix}\chibar_-\,, & \chibar_+\end{pmatrix}}.
\label{eq:doublet}
\end{equation}
With this convention the fields $\psi$ and $\psibar$ satisfy the 
boundary conditions (\ref{eq:bc_rotate}). The 
prefactor has been included in order to obtain the correct normalisation of
the kinetic term, once the action is reduced to the interval $[0,T]$.
In the case at hand the dynamical degrees of freedom are
the fields $\psi(x)$ and $\psibar(x)$ for $0 < x_0<T$ as well
as the non-Dirichlet components at the boundaries, i.e. $\tilde{Q}_-\psi(0,\bfx)$,
$\tilde{Q}_+\psi(T,\bfx)$  and $\psibar(0,\bfx)\tilde{Q}_-$, $\psibar(T,\bfx)\tilde{Q}_+$.
The Dirac operator is implicitly defined when the action is reduced to the time
interval $[0,T]$:
\begin{eqnarray}
   S_f[\chi,\chibar,U] &=&  S_f[\chi_+,\chibar_+,U] +S_f[\chi_-,\chibar_-,U] \nonumber\\
    &=& \frac{1}{2}a^4\!\!\! \sum_{-T < x_0 \le T}\sum_{\bfx}\psibar(x)(D_W+m_0)\psi(x)\nonumber\\
    &=& a^4\!\!\! \sum_{0 \le x_0 \le T}\sum_{\bfx}\psibar(x)({\calD}_W+m_0)\psi(x).
\end{eqnarray}
More explicitly, writing $D_W$ as a time difference operator,
\begin{equation}
  aD_W\psi(x)= -U_0(x)P_-\psi(x+a\hat{\bf 0}) + K\psi(x)
                     - U_0(x-a\hat{\bf 0})^\dagger P_+ \psi(x-a\hat{\bf 0}),
\end{equation}
with the time diagonal kernel being given by
\begin{eqnarray}
 K\psi(x) &=& \left(1+ \frac12\sum_{k=1}^3\left\{ a(\nabla^{}_k+\nabla^\ast
_k)\gamma_k
   -a^2\nabla^\ast_k\nabla^{}_k\right\}\right)\psi(x) \nonumber\\
 &&\mbox{} +\csw \frac{i}{4} a \sum_{\mu,\nu=0}^3\sigma_{\mu\nu}\hat{F}_{\mu\nu}(x)\psi(x),
\end{eqnarray}
one finds
\begin{equation}
   a\calD_W\psi(x) = \begin{cases} \half \tilde{Q}_- K \tilde{Q}_-\psi(x) -\tilde{Q}_-P_-\psi(x+a\hat{0})
                                               & \text{if $x_0=0$,} \\
                              -P_+\tilde{Q}_-\psi(x-a\hat{0})+K\psi(x)-P_-\psi(x+a\hat{0})
                                              & \text{if $x_0=a$},  \\
                                 aD_W\psi(x)  &\text{if $a<x_0<T-a$,}\\
 -P_+\psi(x-a\hat{0}) +K\psi(x) -P_-\tilde{Q}_+\psi(x+a\hat{0})
                                               & \text{if $x_0=T-a$,} \\
                              -\tilde{Q}_+P_+\psi(x-a\hat{0})+\half \tilde{Q}_+K\tilde{Q}_+\psi(x)
                                              & \text{if $x_0=T$},\\
                 \end{cases}.
\label{eq:D_versionI}
\end{equation}
where the temporal links have been set to unit matrices for better readability.
Note that this structure follows from the boundary conditions for both
$\psi$ and $\psibar$, and the explicit factor $1/2$ at the boundaries 
arises from the standard normalisation of the kinetic term. The orbifold 
construction also determines the Sheikholeslami-Wohlert term at the time boundaries.
However, in practical applications it may be easier to set the coefficient of the Sheikholeslami-Wohlert
term at the time boundaries to zero. This amounts to a change of a dimension 5 operator at the boundaries,
i.e.~a change in the action by a term of O($a^2$).

\subsection{Orbifold reflection with an O($a$) offset}

On a hypercubic lattice it is natural to distinguish between site and link reflections.
The orbifold construction just described corresponds to using a site reflection about the
time $x_0=0$. Orbifold constructions based on link reflections correspond to an offset by $a/2$ of the
reflection point, i.e~$x_0=\pm a/2$. Both options may be interesting in practice and are discussed 
in turn, beginning with link reflection about $x_0=-a/2$.
It is convenient to set $T'= T+a$. Then the starting point is the same as in Eq.~({\ref{eq:orb-start}}),
with the replacement $T\rightarrow T'$, 
and the fermion fields are now taken to be anti-periodic with period $2T'$.
The orbifold reflection is defined by
\begin{equation}
  {\cal T}_+: \chi(x) \rightarrow
       i\gamma_0\gamma_5\chi(-a-x_0,{\bf x}),
\qquad
      \chibar(x) \rightarrow
     \chibar(-a-x_0,{\bf x})i\gamma_0\gamma_5.
\end{equation}
and the external gauge field is extended to the interval $[-T',T']$,
\begin{equation}
U_k(-a-x_0,{\bf x})=U_k(x_0,{\bf x}),\qquad U_0(-2a-x_0,{\bf x})^\dagger= U_0(x),
\label{eq:gauge_ext}
\end{equation}
and then $2T'$-periodically continued. This implies that
the spatial gauge fields at the time boundaries are duplicated and
the time-like links connecting the doubled boundary layers are set to unit matrices.

Again the orbifold reflection ${\cal T}_+$ commutes with the Wilson-Dirac operator,
and ${\cal T}_+$-even and -odd quark fields can be defined as previously,
\begin{equation}
   \chi_\pm= \half(1\pm {\cal T}_+)\chi,\qquad \chibar_\pm=\chibar\half(1\pm {\cal T}_+).
\end{equation}
However, it is obvious that these fields will not exactly satisfy the desired
continuum boundary conditions. For instance, for the odd fields one finds,
\begin{equation}
   \chi_-(0,\bfx)= i\gamma_0\gamma_5\chi(-a,\bfx), \qquad 
   \chi_-(T,\bfx)=-i\gamma_0\gamma_5\chi(T+a,\bfx),
\end{equation}
i.e.~boundary conditions would be obtained at $x_0=-a/2$ or $x_0=T+a/2$,
where there are no lattice points available.
One may however say that the boundary conditions are satisfied up to
O($a$) effects. In fact, combining the even and odd fields in quark doublets $\psi$ and
$\psibar$ as in Eq.~(\ref{eq:doublet}), one may write, in terms of the ordinary 
lattice forward and backward derivatives, 
\begin{xalignat}{2}
     \tilde{Q}_+(1-\frac12 a\partial_0^\ast)\psi(x)\vert_{x_0=0} &= 0,
&\qquad \tilde{Q}_-(1+\frac12 a\partial_0)\psi(x)\vert_{x_0=T} &=0, \nonumber\\[0.5ex]
   \psibar(x)\tilde{Q}_+(1-\frac12 a\lvec{\partial}_0^\ast)\vert_{x_0=0} &= 0,
&\qquad \psibar(x) \tilde{Q}_-(1+\frac12 a\lvec{\partial}_0)\vert_{x_0=T} &= 0. 
\label{eq:bc_Oa}
\end{xalignat}
Note that there is no problem with gauge invariance here, as the time-like links
between the doubled boundary layer are set to unit matrices.

Once again the action and the dynamical variables can be reduced 
explicitly to the time interval $[0,T]$, and the final result can be written as
\begin{equation}
   S_f = a^4\sum_{x_0=0}^T\sum_{\bfx} \psibar(x)({\cal D}_W +m_0)\psi(x),
\end{equation}
where the Wilson-Dirac operator is specified by
\begin{equation}
 a{\cal D}_W\psi(x)= \begin{cases}        
         -U_0(x)P_-\psi(x+a\hat{\bf 0}) + (K+i\gamma_5\tau^3P_-)\psi(x)
                                      &    \text{if $x_0=0$,}\\
                     aD_W\psi(x) & \text{if $0<x_0<T$,}\\
        (K+i\gamma_5\tau^3P_+)\psi(x)- U_0(x-a\hat{\bf 0})^\dagger P_+ \psi(x-a\hat{\bf 0})
                                       &    \text{for $x_0=T$.}\\
                      \end{cases}
\label{eq:wilson_dirac}
\end{equation}
The dynamical field variables (i.e.~the integration variables in the functional integral)
are now given by the fields $\psi(x)$ and $\psibar(x)$ for times $0\leq x_0\leq T$.
Note also that the form of the Dirac operator ${\cal D}_W$ can be obtained directly by 
acting with the infinite volume operator $D_W$ on functions $\psi(x)$ for $0\leq x_0\leq T$, 
supplemented with the ``syntactic extension",
\begin{equation}
  \psi(-a,\bfx) = -i\gamma_0\gamma_5\tau^3\psi(0,\bfx),\qquad
  \psi(T+a,\bfx) = i\gamma_0\gamma_5\tau^3\psi(T,\bfx),
\end{equation}
and taken to vanish at all other times. 
Note that a similar statement can not be made for the orbifold construction 
based on the site-reflection ${\cal T}$, as the expression of 
the Dirac operator in Eq.~(\ref{eq:D_versionI}) also refers to the boundary conditions for $\psibar$.

\subsection{Orbifold construction with link reflection about $x_0=a/2$}

We start again from Eq.~({\ref{eq:orb-start}}), but with the replacement $T\rightarrow T'=T-a$.
The fermion fields are taken to be $2T'$-anti-periodic
and the orbifold reflection is defined by
\begin{equation}
  {\cal T}_-: \chi(x) \rightarrow
       i\gamma_0\gamma_5\chi(a-x_0,{\bf x}),
\qquad
      \chibar(x) \rightarrow
     \chibar(a-x_0,{\bf x})i\gamma_0\gamma_5.
\end{equation}
In this case, however, it is not possible to treat the SF gauge field consistently 
as an external field, as this would e.g.~require to set
\begin{equation}
    U_k(a,\bfx)=U_k(0,\bfx),
\end{equation}
i.e.~the gauge field near the boundary would be equal to the gauge field at the boundary.
However, this is not really important as the orbifold construction is no end in itself.
It merely serves to teach us the structure of the Wilson-Dirac operator near the
boundaries for the non-interacting theory. Hence the introduction of gauge fields can be postponed
until the end, and whether or not the interacting theory satisfies the boundary conditions will depend on 
the symmetries and the allowed counterterms.

Proceeding in this way we obtain the action,
\begin{equation}
   S_f = a^4\sum_{x_0=a}^{T-a}\sum_{\bfx} \psibar(x)({\cal D}_W +m_0)\psi(x),
\end{equation}
where the Wilson-Dirac operator is specified by
\begin{equation}
 a{\cal D}_W\psi(x)= \begin{cases}        
         -U_0(x)P_-\psi(x+a\hat{\bf 0}) + (K+i\gamma_5\tau^3P_-)\psi(x)
                                      &    \text{if $x_0=a$,}\\
                     aD_W\psi(x) & \text{if $a<x_0<T-a$,}\\
        (K+i\gamma_5\tau^3P_+)\psi(x)- U_0(x-a\hat{\bf 0})^\dagger P_+ \psi(x-a\hat{\bf 0})
                                       &    \text{for $x_0=T-a$.}\\
                      \end{cases}
\end{equation}
Note that the dynamical field variables in this case are $\psi(x)$ and $\psibar(x)$ at times $0<x_0<T$,
which is analogous to the standard SF for Wilson quarks.

\subsection{The free quark propagator}

In order to check whether the boundary conditions are indeed satisfied in the free theory one
may compute the free quark propagator in a time-momentum representation, 
and compare to its expected continuum limit. There are various ways to proceed and some
explicit results can be found in Appendix~B. In fact the orbifold construction itself allows to 
calculate the propagator, taking the propagator on a hypertorus as starting point. 
The 3 different orbifold procedures can be combined by introducing the parameter $\tau=\pm 1,0$: 
we set $T'=T+\tau a$ and identify the site reflection ${\cal T}$ with ${\cal T}_0$.
Performing  the inverse Fourier transform in time of the propagator in four-momentum space, one obtains
\begin{equation}
S_{\bfp}^{2T'}(x_0-y_0)=\frac{1}{2T'}\sum_{p_0} \rme^{ip_0(x_0-y_0)}\widetilde{D}_W^{-1}(p),
\end{equation}
where 
\begin{equation}
  \widetilde{D}_W(p) = i\tilde\pslash+ m_0+\half a\hat{p}^2,
\end{equation} 
denotes the symbol of the Wilson-Dirac operator with the inverse 
\begin{equation}
 \widetilde{D}_W^{-1}(p) = 
 \frac{-i\tilde\pslash+(m_0+\half a\hat{p}^2)}{\tilde{p}^2+(m_0+\half a\hat{p}^2)^2}.
\end{equation}
Here, lattice momenta have been denoted by
\begin{equation}
  \tilde{p}_\mu= \frac{1}{a}\sin{ap_\mu},\qquad \hat{p}_\mu = \frac{2}{a}\sin\frac{ap_\mu}{2}.
\end{equation}
Note that there are $2 T'/a$ allowed values for $p_0$, parameterised by an integer $k$,
\begin{equation}
   p_0(k)=\frac{(2k+1)\pi}{2T'}, \qquad k=-T'/a,-T'/a+1,\ldots,T'/a,
\end{equation}
so that the sum over $p_0$ can be interpreted as a sum over $k$,
\begin{equation}
\sum_{p_0} \equiv \sum_{p_0(k)}\equiv \sum_{k=-T'/a}^{T'/a-1}.
\end{equation}
All one needs to do to obtain the propagator in the orbifolded setting is to project
on the appropriate ${\cal T}_\tau$-even and -odd field components, 
and take the field normalisation, Eq.~(\ref{eq:doublet}), into account.
Combining the orbifold projection for both flavours by setting
\begin{equation}
    \tilde{\cal T}_\tau = \diag({\cal T}_\tau,-{\cal T}_\tau),
\end{equation}
one obtains  
\begin{eqnarray}
   S^\tau_{\bfp}(x_0,y_0) &=& \frac{1}{2}(1-\tilde{{\cal T}}_\tau)S_{\bfp}^{2T'}(x_0-y_0)(1-\tilde{{\cal T}}_\tau) 
\label{eq:orb1}\\
                     &=&  S_{\bfp}^{2T'}(x_0-y_0)-S_{\bfp}^{2T'}(x_0+y_0+\tau a)i\gamma_0\gamma_5\tau^3.
\label{eq:orb2}
\end{eqnarray}
where the orbifold projection from the left (right) acts on $S_{\bfp}^{2T'}(x_0-y_0)$ taken as a function
of $x_0$ ($y_0$). Note that in going from Eq.~(\ref{eq:orb1}) to Eq.~(\ref{eq:orb2}) the behaviour under 
time reversal has been used, 
\begin{equation}
    S_{\bfp}^{2T'}(t) = i\gamma_0\gamma_5 S_{\bfp}^{2T'}(-t)i\gamma_0\gamma_5.
\end{equation}
It is now straightforward to check that the defining equations of
the fermion propagator are satisfied, i.e.
\begin{equation}
   {\cal D}_W^{\tau} a^3 \sum_{\bfx}\rme^{i\bfp(\bfx-\bfy)}  S^\tau_{\bfp}(x_0,y_0) = a^{-4}\delta_{x,y},
\end{equation}
provided the arguments are restricted to $0\le x_0,y_0\le T$ for $\tau=0,+1$ and to $0 < x_0,y_0 < T$
for $\tau=-1$. Furthermore, the homogeneous boundary conditions are indeed satisfied, for instance,
\begin{equation}
\tilde{Q}_+S^\tau_{\bfp}(0,y_0) = \tilde{Q}_+\left[ S_{\bfp}^{2T'}(-y_0)-S_{\bfp}^{2T'}(-y_0-\tau a)\right],
\end{equation}
vanishes for $\tau=0$ and is of O($a$) for $\tau=\pm 1$. Note that, formally, the boundary conditions are
satisfied at $x_0=-\tau a/2$, which may be used as a check, even though there is no lattice point at $x_0=\pm a/2$.

A more explicit expression for the quark propagator can be obtained using the procedure of
\cite{Luscher:1996vw}. The resulting expressions are given in Appendix~B and coincide
numerically with the above orbifolded expression\footnote{For $\tau=0$, this is correct except for $x_0=y_0=0,T$. 
However, this contact term in the propagator is never referred to provided the boundary quark and anti-quark fields 
are defined at $x_0=a,T-a$ (cf. Sect.~5).}, 
provided one restricts the time arguments to $ 0\leq x_0,y_0 \leq T'$ for $\tau=0,+1$
and to $0< x_0,y_0 \leq T'$ for $\tau=-1$.

Finally, the orbifold procedure makes it clear 
that the free spectrum of $\gamma_5\tau^1 {\cal D}_W^\tau$ is the same as obtained on
a torus with $2T'$-anti-periodic boundary conditions. More precisely, the eigenvalues are of the form
\begin{equation}
   \lambda = \pm\sqrt{\tilde{p}^2+(m_0+\half a\hat{p}^2)^2}.
\end{equation}
In particular, the smallest (in modulus) eigenvalue
is obtained for $\bfp=0$ and is (for $m_0=0$ and $\theta=0$) given by
\begin{equation}
   |\lambda_{\rm min}| = \frac{2}{a}\left|\sin\left(\frac{a\pi}{4 T'}\right)\right| 
   \quad {\buildrel{a\rightarrow 0}\over\sim} \quad \frac{\pi}{2T},
\end{equation}
which converges to the known continuum result~\cite{Sint:1993un}. 

\subsection{Orbifold construction, summary}

The orbifold construction yields the structure of the Wilson-Dirac operator
near the time boundaries, such that the chirally rotated boundary conditions are realised
in the free theory, possibly up to O($a$) effects. 
Using the link reflections  ${\cal T}_\pm$, the action of the 
Wilson-Dirac operator on the interval can be specified by a syntactic extension 
of the function space it acts upon. This is not the case for the site reflection ${\cal T}$,
where in addition one needs to refer to the boundary condition for the anti-quark fields.

In all three cases the relation,
\be
   (\gamma_5\tau^1 \calD_W)^\dagger = \gamma_5\tau^1 \calD_W,
\ee
implies that the determinant of the Wilson Dirac operator is real.
Moreover, a well defined eigenvalue problem is obtained for $\gamma_5\tau^1 \calD_W$. 
In the free theory, the eigenvalues are determined by the possible values of
the time momentum component for anti-periodic boundary conditions 
on the doubled time interval. In particular, the lowest eigenvalue is determined by
$\pi/(2T')$, where $T'=T+\tau a$ and $\tau =\pm 1,0$ for the three versions of the orbifold discussed
above.

Concerning mass terms, we have seen that the standard mass term can be added without
any problem. Adding a twisted mass term $i\muq\gamma_5\tau^3$ is, on the other hand, only 
possible by including a time profile following ref.~\cite{Taniguchi:2004gf}. 
Introducing the mass term in another flavour direction, 
e.g. $i\muq\gamma_5\tau^2$ is however possible without such a time
profile, and corresponds to a twisted mass term in the standard SF basis, too. In order to
incorporate such a mass term in the orbifold construction it is best to start
directly with flavour doublet fields and define the orbifold reflection directly
for this case. The twisted mass term is then seen to commute with 
the spin-flavour matrix $i\gamma_0\gamma_5\tau^3$ in the orbifold 
reflection $\tilde{\cal T}_\tau$, and therefore with $\tilde{\cal T}_\tau$ itself.

%% file: section5.tex
\section{Symmetries, renormalisation and  O($a$) improvement}

Given the fermionic lattice action with the correct boundary conditions at
tree level, the next step consists in identifying the exact lattice symmetries and the allowed boundary
counterterms. One then needs to choose a counterterm basis and discuss their lattice implementation
and their effect on the basic fermionic 2-point functions.

\subsection{Counterterms of dimension 3 and 4}

The Ward identities, Eqs.~(\ref{eq:ward1},\ref{eq:ward2}) suggest that the symmetry 
breaking induced by the chirally rotated  SF boundary conditions is quite similar 
to the breaking by a twisted mass term, except that the
breaking terms are localised at the boundaries and come with dimensionless coefficients. 
Therefore it should not be too surprising that the symmetries of the lattice action 
in the chirally rotated Schr\"odinger functional are the same as in twisted mass 
lattice QCD~\cite{Frezzotti:2000nk}, except for those space-time symmetries 
which mix time and spatial directions. In particular, there is charge conjugation, spatial lattice rotations, 
space and time reflections  combined with a flavour exchange and global U(1) vector like rotations with generator $\tau^3/2$. 

Given the symmetries, a list of fermionic operators of dimension 3 and 4 can
be worked out. Their integrals over space, taken at $x_0=0$ and $x_0=T$ 
define the possible counterterms to the lattice action that are needed to renormalise and O($a$) improve
the boundary effects in the Schr\"odinger functional. We do not need to discuss the 
bulk counterterms here, as these are the same as in infinite volume~\cite{Luscher:1996sc}.

Symmetrising with respect to charge conjugation we find the following operators 
allowed by the symmetries: there are 3 operators of mass dimension 3,
\begin{eqnarray}
  \op{1} &=& \psibar\gamma_0 \tilde{Q}_+\psi 
          - \psibar \gamma_0 \tilde{Q}_-\psi = \psibar i\gamma_5\tau^3\psi,
\label{eq:counter1}\\
  \op{2} &=& \psibar \tilde{Q}_+\psi,\\
  \op{3} &=& \psibar \tilde{Q}_-\psi,
\end{eqnarray}
and further 8 operators of mass dimension 4:
\begin{eqnarray}
  \op{4} &=& \psibar\,\tilde{Q}_{+}\gamma_k D_k\psi
             -\psibar\,\lvec{D}_k\gamma_k \tilde{Q}_{+}\psi,\\
  \op{5} &=& \psibar\,\tilde{Q}_{-}\gamma_k D_k\psi
             -\psibar\,\lvec{D}_k\gamma_k \tilde{Q}_{-}\psi,\\
  \op{6} &=& \psibar\,\tilde{Q}_{+}\gamma_0 D_0\psi
             -\psibar\,\lvec{D}_0\gamma_0 \tilde{Q}_{+}\psi,\\
  \op{7} &=& \psibar\,\tilde{Q}_{-}\gamma_0 D_0\psi
             -\psibar\,\lvec{D}_0\gamma_0 \tilde{Q}_{-}\psi,\\
  \op{8} &=& \psibar\,\tilde{Q}_{+} D_0\psi
             +\psibar\,\lvec{D}_0 \tilde{Q}_{+}\psi,\\
  \op{9} &=& \psibar\,\tilde{Q}_{-} D_0\psi
             +\psibar\,\lvec{D}_0 \tilde{Q}_{-}\psi,  \\
  \op{10}&=& \psibar\,\tilde{Q}_{+}\gamma_0\gamma_k D_k\psi
             +\psibar\,\lvec{D}_k\gamma_k \gamma_0 \tilde{Q}_{+}\psi,  \\
  \op{11}&=& \psibar\,\tilde{Q}_{-}\gamma_0\gamma_k D_k\psi
             +\psibar\,\lvec{D}_k\gamma_k \gamma_0\tilde{Q}_{-}\psi.
\label{eq:counter11}
\end{eqnarray}
The dimension 4 operators are related by the equations of motion 
\begin{equation}
   \Dslash\psi=0,\qquad  \psibar \lvec\Dslash =0,
\end{equation}
which imply four relations
\begin{eqnarray}
  \op{4}+\op{6}  &=& 0,\\
  \op{5}+\op{7}  &=& 0,\\
  \op{8}+\op{10} &=& 0,\\
  \op{9}+\op{11} &=& 0.
\end{eqnarray}
Furthermore, the combination,
\begin{equation}
   \op{10}-\op{11}= \partial_k\left(\psibar\,\gamma_k i\gamma_5\tau^3\psi \right),
\end{equation}
yields a total spatial derivative and therefore does not contribute to the action.
In summary, we have 5 relations among the 8 counterterms of mass dimension 4.
No such simplification is possible for the dimension 3 operators. Hence one
needs to consider 3 O(1) and 3 O($a$) counterterms, which must be added to the
SF lattice action with appropriately chosen coefficients. 

\subsection{Formal continuum theory with inhomogeneous boundary conditions}

In order to understand how the boundary counterterms affect SF correlation functions 
one would ideally start from the lattice theory with {\em inhomogeneous} boundary conditions,
following the steps taken for the standard SF~\cite{Luscher:1996vw}. Given the
counterterms to the action, the renormalised and O($a$) improved 2-point functions are obtained 
by differentiating  with respect to the fermionic boundary values and bulk source fields,
which are subsequently set to zero. Arbitrary SF correlation functions can be built from these 
basic 2-point functions, and their renormalization and O($a$) improvement properties then follow, too.
The main problem one faces is that the formulation of the lattice theory with 
inhomogeneous boundary conditions is not directly available. Indeed,
our construction of the lattice theory based on the orbifold procedure yields
homogeneous boundary conditions. While this is all that will be needed eventually,
some alternative method is required to understand the r\^ole of the boundary counterterms.
The approach taken here is based on the fact that the allowed
counterterms of dimension 3 and 4 completely fix the structure of the unknown 
lattice theory to O($a$). This turns out to be sufficient to extract the desired information 
which may then be used to parametrize the renormalized lattice theory.

As a first step let us go through the formal continuum theory with inhomogenous boundary conditions, 
specified by
\begin{xalignat}{2}
     \tilde{Q}_+\psi(x)\vert_{x_0=0} &=\rho(\bfx),    
    &\tilde{Q}_-\psi(x)\vert_{x_0=T} &=\rho'(\bfx),\nonumber\\
   \psibar(x)\tilde{Q}_+\vert_{x_0=0} &= \rhobar(\bfx),   
    &\psibar(x)\tilde{Q}_-\vert_{x_0=T} &= \rhobar'(\bfx). 
\label{eq:bc_inhomogeneous}
\end{xalignat}
As in the case of the standard SF~\cite{Sint:1993un}, the action is not just given by the 
bulk action but also requires local boundary terms. We thus make the ansatz
\begin{eqnarray}
  S_f^{\rm cont}[\psi,\psibar,A] &=&\int\rmd^4 x\,\psibar(x)\half\lrvec{\Dslash} \psi(x) 
  + \int \rmd^3\bfx\, \left[\psibar(x) R_0(\bfx)\psi(x)\right]_{x_0=0}\nonumber\\
  &&  + \int \rmd^3\bfx\, \left[\psibar(x) R_T(\bfx) \psi(x)\right]_{x_0=T},
\end{eqnarray}
with $R_0(\bfx)$ and $R_T(\bfx)$ to be determined.
Here, $A_\mu$ denotes the external gauge field, and the symmetrisation of the derivative 
implies that bulk and boundary parts of the action density must be separately invariant under charge conjugation. 
In order to determine the boundary terms we now follow the steps in ref.~\cite{Sint:1993un}, and demand
that the action must have smooth stationary points, 
i.e.~classical solutions $\psi_{\rm cl}$ and $\psibar_{\rm cl}$
which satisfy both the boundary conditions, Eqs.~(\ref{eq:bc_inhomogeneous}}), 
and the equations of motion,
\begin{equation}
   \Dslash \psi_{\rm cl} =0,\qquad 
   \psibar_{\rm cl}\lvec{\Dslash} =0.
\end{equation}
This is ensured if $R_0$ and $R_T$ are of the form
\begin{equation}
  R_0(\bfx) = -\frac{i}{2}\gamma_5\tau^3+\tilde{Q}_+\Gamma_0(\bfx)\tilde{Q}_+ \,,\qquad
  R_T(\bfx) = -\frac{i}{2}\gamma_5\tau^3+\tilde{Q}_-\Gamma_T(\bfx)\tilde{Q}_- ,
\label{eq:R0RT}
\end{equation}
with arbitrary $\Gamma_{0}$ and $\Gamma_{T}$, subject only to symmetry requirements. 
Imposing the continuum symmetries  of the standard SF, in particular parity and flavour symmetries,
one finds that these terms must vanish, so that,
\begin{equation}
  R_0(\bfx) = R_T(\bfx) = -\frac{i}{2}\gamma_5\tau^3.
\end{equation}
Not surprisingly, the resulting action coincides with the chirally rotated 
continuum action of the standard SF~\cite{Sint:1993un}.
Using this action, we now consider the functional integral over the fermion fields,
\begin{equation}
  {\cal Z}_F[\rho,\rhobar;\rhoprime,\rhobarprime;\eta,\etabar] = 
  \int D[\psi,\psibar]\exp\left\{-S_f^{\rm cont}[\psi,\psibar,A] + (\etabar,\psi) + (\psibar,\eta)\right\},
\end{equation}
which implicitly depends on the fermionic boundary values in Eq.~(\ref{eq:bc_inhomogeneous}).
We have also included source fields $\eta$ and $\etabar$ in the bulk using the notation
\begin{equation}
 (\etabar,\psi) = \int\rmd^4 x\, \etabar(x) \psi(x).
\end{equation}
The functional integration is best carried out after shifting variables,
\begin{equation}
   \psi=\psi_{\rm cl}+v,\qquad \psibar=\psibar_{\rm cl}+\bar{v},
\end{equation}
where the fluctuation fields $v$ and $\bar{v}$ satisfy homogeneous boundary conditions,
and the classical solutions of the field equations are explicitly given by
\begin{eqnarray}
\psi_{\rm cl}(x) &=& \int\rmd^3\bfy \left[ S(x;0,\bfy)\gamma_0 \tilde{Q}_+\rho(\bfy) -
                                          S(x;T,\bfy)\gamma_0 \tilde{Q}_-\rhoprime(\bfy)\right],
\label{eq:psicl}\\
\psibar_{\rm cl}(x) &=& \int\rmd^3\bfy \left[\rhobarprime(\bfy)\gamma_0 \tilde{Q}_+ S(T,\bfy;x) -
                                          \rhobar(\bfy)\gamma_0 \tilde{Q}_- S(0,\bfy;x)\right].
\label{eq:psibarcl}
\end{eqnarray}
Here, $S(x,y)$ denotes the continuum quark propagator in the external gauge field $A_\mu$. 
Integration over the $v$ and $\bar{v}$ yields  
\begin{equation}
\ln {\cal Z}_F = \ln {\cal Z}_{F,0} 
- S_f^{\rm cont}[\psi_{\rm cl},\psibar_{\rm cl},A]+ (\etabar,S\eta) 
 + (\etabar,\psi_{\rm cl}) + (\psibar_{\rm cl},\eta),
\label{Z_F}
\end{equation}
with
\begin{equation}
 {\cal Z}_{F,0} = {\cal Z}_{F}[0,0;0,0;0,0],\qquad
 (S\eta)(x) = \int\rmd^4y\, S(x,y)\eta(y).   
\end{equation}
The dependence upon the source fields becomes completely explicit when inserting the 
classical solutions (\ref{eq:psicl}) and (\ref{eq:psibarcl}) into equation~(\ref{Z_F}).
In particular the continuum action of the classical fields takes the form
\begin{eqnarray}
S_f^{\rm cont}[\psi_{\rm cl},\psibar_{\rm cl},A] &=& 
\int\rmd^3{\bfx}\,\rmd^3{\bfy}\,\Bigl[
                                 \rhobar(\bfx)\gamma_0\tilde{Q}_-S(0,\bfx;0,\bfy)\tilde{Q}_-\gamma_0\rho(\bfy)\nonumber\\
 && \hphantom{2\int\rmd^301} - \rhobar(\bfx)\gamma_0\tilde{Q}_-S(0,\bfx;T,\bfy)\tilde{Q}_+\gamma_0\rhoprime(\bfy)\nonumber\\
 && \hphantom{2\int\rmd^301} - \rhobarprime(\bfx)\gamma_0\tilde{Q}_+S(T,\bfx;0,\bfy)\tilde{Q}_-\gamma_0\rho(\bfy)\nonumber\\
 && \hphantom{2\int\rmd^301} + \rhobarprime(\bfx)\gamma_0\tilde{Q}_+S(T,\bfx;T,\bfy)\tilde{Q}_+\gamma_0\rhoprime(\bfy)\Bigr]\,.
\label{eq:Scont}
\end{eqnarray}
It is convenient to follow ref.~\cite{Luscher:1996vw} and define fermionic expectation values
through\footnote{The definition is such that the usual Euclidean expectation value is obtained by a subsequent integration over
the gauge field, i.e.~$\langle O \rangle = \langle  [O]_F \rangle_G$, where the effective gauge field measure in $\langle \cdot\rangle_G$ 
includes both the pure gauge action and the fermionic determinant.}
\begin{equation}
   [O]_F = \left\{\frac{1}{{\cal Z}_F}O{\cal Z}_F\right\}_{\rhobar=\cdots=\eta=0}.
\end{equation}
Here the observables $O$ may contain derivatives with respect to the source fields. 
Since the effective action is bilinear in the source fields, the fermionic expectation value of 
any observable can be given in terms of the basic fermionic 2-point functions,
defined as all possible second derivatives of $\ln {\cal Z}_F$ with respect to the source fields. 
In order to be compatible with Eqs.~(\ref{eq:zeta_alpha}),
for $\alpha=\pi/2$, one identifies
\begin{xalignat}{2}
   \zeta(\bfx) & \leftrightarrow -\frac{\delta}{\delta[\rhobar(\bfx)\gamma_0]},    
    & \zetabar(\bfx)   &\leftrightarrow -\frac{\delta}{\delta[\gamma_0\rho(\bfx)]}, \nonumber\\
   \zetaprime(\bfx) &\leftrightarrow \frac{\delta}{\delta[\rhobarprime(\bfx)\gamma_0]}  , 
    &  \zetabarprime(\bfx)   &\leftrightarrow \frac{\delta}{\delta[\gamma_0\rhoprime(\bfx)]}. 
\label{eq:zetafunc}
\end{xalignat}
Using this notation, and, similarly,
\begin{equation}
   \psi(x)\leftrightarrow \frac{\delta}{\delta\etabar(x)},\qquad 
   \psibar(x)\leftrightarrow -\frac{\delta}{\delta\eta(x)},
\label{eq:zetafunc1}
\end{equation}
all the basic fermionic 2-point functions are given by
\begin{eqnarray}
 [\psi(x)\psibar(y)]_F &=& S(x,y), \label{eq:2pt1}\\[0ex]
 [\psi(x)\zetabar(\bfy)]_F &=& S(x;0,\bfy)\tilde{Q}_-,\label{eq:2pt2}\\[0ex]
 [\psi(x)\zetabarprime(\bfy)]_F &=& S(x;T,\bfy)\tilde{Q}_+,\\[0ex]
 [\zeta(\bfx)\psibar(y)]_F &=& \tilde{Q}_-S(0,\bfx;y),\\[0ex]
 [\zetaprime(\bfx)\psibar(y)]_F &=& \tilde{Q}_+S(T,\bfx;y),\\[0ex]
 [\zeta(\bfx)\zetabar(\bfy)]_F &=& \tilde{Q}_-S(0,\bfx;0,\bfy)\tilde{Q}_-,\label{eq:2pt6}\\[0ex]
 [\zeta(\bfx)\zetabarprime(\bfy)]_F &=& \tilde{Q}_-S(0,\bfx;T,\bfy)\tilde{Q}_+,\\[0ex]
 [\zetaprime(\bfx)\zetabar(\bfy)]_F &=& \tilde{Q}_+S(T,\bfx;0,\bfy)\tilde{Q}_-,\\[0ex]
 [\zetaprime(\bfx)\zetabarprime(\bfy)]_F &=& \tilde{Q}_+S(T,\bfx;T,\bfy)\tilde{Q}_+. \label{eq:2pt9}
\end{eqnarray}
Thus, in the formal continuum theory, the insertion into Eqs.~(\ref{eq:2pt2}--\ref{eq:2pt9}) 
of the non-Dirichlet field components at the time boundaries~(\ref{eq:zeta_alpha}), 
is indeed equivalent to their inclusion as functional derivatives, Eqs.~(\ref{eq:zetafunc},\ref{eq:zetafunc1}).

\subsection{Including the counterterm action}

The correctly renormalised and O($a$) improved lattice theory is expected to converge to the formal continuum 
theory up to terms of O($a^2$). By adding the counterterms with some as yet undetermined coefficients
one essentially defines Symanzik's effective action for a generic lattice theory the formulation of 
which is not directly available. Obviously, the inclusion of dimension 3 boundary counterterms 
with arbitrary coefficients may alter the very continuum theory including the boundary conditions. 
This point will be addressed in Sect.~6, but we will ignore it here, as we are 
only looking for a structural result: by keeping arbitrary coefficients 
we may monitor their effect on the Dirac operator and the basic fermionic 2-point functions. 
This will then enable us to parameterize the counterterms in the lattice formulation. 

For definiteness, we use the equations of motion to obtain a counterterm basis for the massless theory,
where all operators with temporal derivatives are eliminated. The resulting set of operators,
\begin{equation}
  \op{1-5}, \op{10},
\label{eq:counterD}
\end{equation}
will be used at both time boundaries. 
The boundary counterterm action is given by
\begin{equation}
 S_{f,{\rm b}}^{\rm c.t.} = \sum_{i=1}^3 c_i S_{f,{\rm b}}^{(i)} + \sum_{i=4}^5 c_i a S_{f,{\rm b}}^{(i)} 
 +  c_{10} a S_{f,{\rm b}}^{(10)},
\end{equation}
where the indiviual terms are specified by
\begin{eqnarray}
S_{f,{\rm b}}^{(1)} &=& \int\rmd^3\bfx\left(\op{1}\vert_{x_0=0} + \op{1}\vert_{x_0=T}\right),\\
S_{f,{\rm b}}^{(2)} &=& \int\rmd^3\bfx\left(\op{2}\vert_{x_0=0} + \op{3}\vert_{x_0=T}\right),\\
S_{f,{\rm b}}^{(3)} &=& \int\rmd^3\bfx\left(\op{3}\vert_{x_0=0} + \op{2}\vert_{x_0=T}\right),\\
S_{f,{\rm b}}^{(4)} &=& \int\rmd^3\bfx\left(\op{4}\vert_{x_0=0} + \op{5}\vert_{x_0=T}\right),\\
S_{f,{\rm b}}^{(5)} &=& \int\rmd^3\bfx\left(\op{5}\vert_{x_0=0} + \op{4}\vert_{x_0=T}\right),\\
S_{f,{\rm b}}^{(10)} &=& \int\rmd^3\bfx\left(\op{10}\vert_{x_0=0} - \op{10}\vert_{x_0=T}\right),
\end{eqnarray}
taking into account the behaviour under time reflections.
Including the counterterm action, the total fermionic action becomes
\begin{equation}
 S_f^{\rm total} = S_f^{\rm cont} + S_{f,{\rm b}}^{\rm c.t.} + a S_1.
 \label{eq:S_tot}
\end{equation}
The last term contains the volume O($a$) counterterms [cf. Eq.~{\ref{eq:S1})].
Note that the elimination of counterterms containing time derivatives implies
that O($a$) effects such as in Eqs.~(\ref{eq:bc_Oa}) are absent. 
Hence, disregarding O($a^2$) effects, the boundary conditions can still be imposed at $x_0=0$ and $x_0=T$.
In a first attempt we will thus try to maintain the boundary 
conditions~(\ref{eq:bc_inhomogeneous}) to O($a$). The first question to answer is whether there
are still smooth classical solutions with the O($a$) enhanced action. Varying the action 
one immediately notices that there are three categories of counterterms. First, there are those
which only refer to field components at the boundaries which are subject to the Dirichlet
conditions, such as $\tilde{Q}_+\psi$ and $\psibar\tilde{Q}_+$ at $x_0=0$. This is the
case for the counterterms $\propto c_2,c_4$. Then there are those which mix Dirichlet
and non-Dirichlet components ($\propto c_1,c_{10}$) and finally the ones only referring to non-Dirichlet
components ($\propto c_3,c_5$). Since the non-Dirichlet components are integration variables in the
functional integral, the last category of counterterms can be absorbed in a re-definition of 
the Dirac operator. The same holds true for the bulk O($a$) improvement terms in $S_1$. 
This leads to a redefined Dirac operator which we shall denote by $\Dslash^{(r)}$. 
Defining classical solutions,
\begin{equation}
   \Dslash^{(r)} \psi_{\rm cl} =0,\qquad 
   \psibar_{\rm cl}\lvec{\Dslash}^{(r)} =0,
\end{equation}
subject to the boundary conditions~(\ref{eq:bc_inhomogeneous}),
the question is whether the presence of the counterterms $\propto c_1,c_2,c_4$ and $c_{10}$, 
allows for the classical solutions to be smooth.
We  need to check what kind of structures $R_0$ and $R_T$ correspond
to the counterterms $\propto c_1,c_2,c_4$ and $c_{10}$, and whether these take the form
given in Eq.~(\ref{eq:R0RT}). This is indeed the case for the counterterms $\propto c_2,c_4$, 
which induce non-vanishing $\Gamma_0$ and $\Gamma_T$ terms. 
However, the counterterms $\propto c_1,c_{10}$ cannot be cast in this form. 
Some thought reveals that one may take the counterterms $\propto c_1,c_{10}$ 
into consideration by a modification 
of the boundary conditions rather than by including them in the action. 
We set 
\begin{xalignat}{2}
     \tilde{Q}_+\psi(x)\vert_{x_0=0} &=\rho_r(\bfx),    
    &\tilde{Q}_-\psi(x)\vert_{x_0=T} &=\rhoprime_r(\bfx),\nonumber\\
   \psibar(x)\tilde{Q}_+\vert_{x_0=0} &= \rhobar_r(\bfx),   
    &\psibar(x)\tilde{Q}_-\vert_{x_0=T} &= \rhobarprime_r(\bfx), 
\label{eq:bc_inhomogeneous_r}
\end{xalignat}
where we have re-defined the boundary values as follows,
\begin{eqnarray}
    \rho_r(\bfx) &=& \sqrt{1-2c_1}\left(1 - \bar{c}_{10}a\gamma_k D_k\right)\rho(\bfx),\\
    \rhobar_r(\bfx) &=& \rhobar(\bfx)\left(1 + \bar{c}_{10}a\gamma_k \lvec{D}_k\right)\sqrt{1-2c_1},
    \qquad \bar{c}_{10}= c_{10}/(1-2c_1),
    \label{eq:rhor}
\end{eqnarray}
and analogously for the primed fields. When considered as a function of  
$\rho, \rhobar, \rhoprime$  and $\rhobarprime$ the reduced action
\begin{equation}
 S_f^{(r)} = S_f^{\rm cont} + \sum_{i=2}^3 c_i S_{f,{\rm b}}^{(i)} + \sum_{i=4}^5 c_i a S_{f,{\rm b}}^{(i)} + a S_1.
\end{equation}
coincides up to terms of O($a^2$) with the total action, Eq.~(\ref{eq:S_tot}),
taken with the original boundary conditions.
We re-define the Dirac operator as before, so that the fermionic action takes the form
\begin{eqnarray}
  S_f^{(r)}[\psi,\psibar,A] &=&\int\rmd^4 x\,\psibar(x)\half\lrvec{\Dslash^{(r)}} \psi(x) 
 -\half \int \rmd^3\bfx\, \left[\psibar(x) i\gamma_5\tau^3 \psi(x)\right]_{x_0=0}\nonumber\\
         &&  -\half \int \rmd^3\bfx\, \left[\psibar(x) i\gamma_5\tau^3 \psi(x)\right]_{x_0=T}\nonumber \\
	 &&  +\, c_2\, S_{f,{\rm b}}^{(2)}[\psi,\psibar,A]  + c_4\, a S_{f,{\rm b}}^{(4)}[\psi,\psibar,A]. 
  \label{eq:S_f}
\end{eqnarray}
The classical solutions $\psi_{{\rm cl},r}$ 
and $\psibar_{{\rm cl},r}$ are defined by
\begin{equation}
   \Dslash^{(r)} \psi_{{\rm cl},r} =0,\qquad 
   \psibar_{{\rm cl},r}\lvec{\Dslash}^{(r)} =0,
\end{equation}
subject to the modified boundary conditions Eq.~(\ref{eq:bc_inhomogeneous_r}).
Explicit expressions  are similar to Eqs.~(\ref{eq:psicl},\ref{eq:psibarcl}),
where the differences are accounted for by replacing the boundary values $\rho\rightarrow \rho_r$ etc.,
and the propagator $S(x,y)\rightarrow S_r(x,y)$, defined as the Green's function for $\Dslash^{(r)}$.
Separating classical and fluctuation fields the fermionic action splits in two parts,
\begin{equation}
 S_f^{(r)}[\psi_{{\rm cl},r}+v, \psibar_{{\rm cl},r}+\bar{v},A] = 
 S^{(r)}_f[v,\bar{v},A] + S^{(r)}_f[\psi_{{\rm cl},r},\psibar_{{\rm cl},r},A].
\end{equation}
The second term on the r.h.s.~contains the whole dependence on the boundary source fields,
and can be written in the form,
\begin{eqnarray}
S^{(r)}_f[\psi_{{\rm cl},r},\psibar_{{\rm cl},r},A] &=&
c_2\int\rmd^3{\bfx}\Bigl[\rhobar_r(\bfx)\tilde{Q}_+\rho_r(\bfx) 
     + \rhobarprime_r(\bfx)\tilde{Q}_-\rhoprime_r(\bfx)\Bigr]\nonumber\\
&&\mbox{+}\, 2 c_4 \,a \int\rmd^3{\bfx}\Bigl[\rhobar_r(\bfx)\tilde{Q}_+\gamma_kD_k\rho_r(\bfx) 
     + \rhobarprime_r(\bfx)\tilde{Q}_- \gamma_k D_k\rhoprime_r(\bfx)\Bigr]\nonumber\\
&&\mbox{+} \int\!\rmd^3{\bfx}\rmd^3{\bfy}\,\Bigl[
    \rhobar_r(\bfx)\gamma_0\tilde{Q}_-S_r(0,\bfx;0,\bfy)\tilde{Q}_-\gamma_0\rho_r(\bfy)\nonumber\\
 && \hphantom{\int\rmd^3{\bfx}012} - \rhobar_r(\bfx)\gamma_0\tilde{Q}_-S_r(0,\bfx;T,\bfy)\tilde{Q}_+\gamma_0\rhoprime_r(\bfy)\nonumber\\
 && \hphantom{\int\rmd^3{\bfx}012} - \rhobarprime_r(\bfx)\gamma_0\tilde{Q}_+S_r(T,\bfx;0,\bfy)\tilde{Q}_-\gamma_0\rho_r(\bfy)\nonumber\\
 && \hphantom{\int\rmd^3{\bfx}012} + \rhobarprime_r(\bfx)\gamma_0\tilde{Q}_+S_r(T,\bfx;T,\bfy)\tilde{Q}_+\gamma_0\rhoprime_r(\bfy)\Bigr].
\end{eqnarray}
Introducing again the bulk source fields, we arrive at the expression of the 
fermionic generating functional,
\begin{eqnarray}
\ln {\cal Z}^{(r)}_F &=& \ln {\cal Z}^{(r)}_{F,0}  - S^{(r)}_f[\psi_{{\rm cl},r},\psibar_{{\rm cl},r},A]+(\etabar,S_r\eta)\nonumber\\
               && +\,(\etabar,\psi_{{\rm cl},r})+ (\psibar_{{\rm cl},r},\eta),
\end{eqnarray}
where the dependence on the source fields is completely explicit.
Before taking derivatives with respect to the source fields one may partially integrate the terms proportional 
to $\bar{c}_{10}$ so that all the spatial derivatives act on the fermion propagator.
It is then straightforward to compute all fermionic 2-point functions, e.g.
\begin{eqnarray}
  [\zeta(\bfx)\psibar(y)]_F &=&  \sqrt{1-2c_1} \left(1 - \bar{c}_{10}a\gamma_k D_k\right)\tilde{Q}_- S_r(0,\bfx;y),\\[0ex]
  [\psi(x)\zetabar(\bfy)]_F &=&   \sqrt{1-2c_1} S_r(x;0,\bfy)\tilde{Q}_-\left(1+\bar{c}_{10}a\gamma_k\lvec{D}_k\right).
\end{eqnarray}
By comparing with Eqs.~(\ref{eq:2pt1}--\ref{eq:2pt9}), we thus obtain 
a clear picture of how the counterterms affect the basic 2-point functions and
the Dirac operator.  Based on this information one may now 
parametrize both the O($a$) improved Wilson-Dirac operator and the basic 2-point functions on the lattice.

\subsection{Lattice parameterisation}

We here focus primarily on the version of the $\chi$SF obtained with the 
orbifold reflection ${\cal T}_\tau$ with $\tau=+1$ (cf.~Sect.~4). 
Hence, the fermionic fields at Euclidean times 
$x_0=0$ and $x_0=T$ are integration variables in the functional integral 
and the counterterms in the chosen basis can be defined 
at $x_0=0$ and $x_0=T$. Starting with the Wilson-Dirac operator, the necessary counterterms are included by replacing 
${\cal D}_W$ of Eq.~(\ref{eq:wilson_dirac}) with ${\cal D}_W +\delta {\cal D}_W$, where
we define
\begin{eqnarray}
 \delta {\cal D}_W \psi (x)  &=& \left(\delta_{x_0,0}+\delta_{x_0,T}\right)
\Bigl[\left(z_f-1\right)+ \left(d_s-1\right) a{\bf D}_s\Bigr]\psi(x),
\label{eq:deltaDW}\\
 {\bf D}_s &=& \frac12 (\nabla^{}_k+\nabla^\ast_k)\gamma_k.
 \label{eq:bfDs}
\end{eqnarray}
On the lattice the fields are always subject to (a lattice version of) 
homogeneous boundary conditions (cf. subsection 4.2), so that $\psi$ and $\psibar$ correspond 
to the fluctuation fields $\bar{v}$ and $v$ in the continuum theory. 
This implies relations such as
\begin{equation}
   \Bigl.\psibar(x)\psi(x)\Bigr\vert_{x_0=0} = \Bigl.\psibar(x)\tilde{Q}_- \psi(x)\Bigr\vert_{x_0=0} + \rmO(a)
\end{equation}
which have been used to simplify the parametrization\footnote{When defining
$\delta {\cal D}_W$ for $\tau=0$ one needs to keep the projectors $\tilde{Q}_\pm$, in order to project
out the Dirichlet component of the fields at $x_0=0$ and $x_0=T$ which are not integrated over in the
functional integral~(cf.~Sect.~4). In the case $\tau=-1$, one may use the same parameterisation as in Eq.~(\ref{eq:deltaDW}) 
except that the Kronecker $\delta$'s should be localised at $x_0=a$ and $x_0=T-a$.}~(\ref{eq:deltaDW}).
Note that O($a$) corrections to such continuum relations either re-define
the coefficients of the existing O($a$) counterterms or modify O($a^2$) terms in the action, 
which are considered irrelevant in the present context. 

There is also considerable freedom in parametrizing the basic 2-point functions.
We will here follow the convention used in the standard SF  and obtain fermionic boundary 
fields from the fermion fields at times $x_0=a$ and $x_0=T-a$, parallel transported back 
to the boundaries at $x_0=0$ and $x_0=T$. In view of this, 
we introduce a shorthand notation for the parallel transported boundary-to-boundary propagators,
\begin{eqnarray}
   \bar{S}(0,\bfx;0,\bfy) &=& U_0(0,\bfx)S(a,\bfx;a,\bfy)U_0(0,\bfy)^\dagger,\\
   \bar{S}(0,\bfx;T,\bfy) &=& U_0(0,\bfx)S(a,\bfx;T-a,\bfy)U_0(T-a,\bfy)\\
   \bar{S}(T,\bfx;0,\bfy) &=& U_0(T-a,\bfx)^\dagger S(T-a,\bfx;a,\bfy)U_0(0,\bfy)^\dagger,\\
   \bar{S}(T,\bfx;T,\bfy) &=&  U_0(T-a,\bfx)^\dagger S(T-a,\bfx;T-a,\bfy)U_0(T-a,\bfy).
\end{eqnarray}
where the quark propagator is defined by
\begin{equation}
  \left({\cal D}_W +\delta {\cal D}_W + m_0\right) S(x,y)= a^{-4}\delta_{x,y},
\end{equation}
and we set $m_0=\mcr$.
The unrenormalised, O($a$) improved fermionic 2-point functions in the massless lattice theory
are then parametrized as follows:
\begin{eqnarray}
[\psi(x) \psibar(y)]_F &=& S(x,y)\,, \\[0ex]
[\psi(x) \zetabar(\bfy)]_F &=& 
 S(x;a,\bfy)U_0(0,\bfy)^\dagger\left(1-\bar{d}_s a\lvec{{\bf D}}_s\right)\tilde{Q}_-\,,\\[0ex]
[\psi(x) \zetabarprime(\bfy)]_F &=& 
S(x;T-a,\bfy)U_0(T-a,\bfy)\left(1-\bar{d}_s a\lvec{{\bf D}}_s\right)\tilde{Q}_+\,,\\[0ex]
[\zeta(\bfx) \psibar(y)]_F &=& 
\tilde{Q}_-\left(1+ \bar{d}_s a{\bf D}_s \right) U_0(0,\bfx)S(a,\bfx;y)\,, \\[0ex]
[\zetaprime(\bfx) \psibar(y)]_F &=&  
\tilde{Q}_+\left(1+ \bar{d}_s a{\bf D}_s \right) U_0(T-a,\bfx)^\dagger S(T-a,\bfx;y)\,,\\[0ex]
[\zeta(\bfx) \zetabar(\bfy)]_F &=&  
\tilde{Q}_-\left(1+ \bar{d}_s a{\bf D}_s \right)\Bigl[\bar{S}(0,\bfx;0,\bfy) \nonumber\\
      &&  - (\tilde{z}_f +\tilde{d}_s a{\bf D}_s)\delta(\bfx-\bfy)\Bigr]
        \left(1-\bar{d}_s a\lvec{{\bf D}}_s\right)\tilde{Q}_-\,,
\label{eq:contact1}\\[0ex]
[\zeta(\bfx) \zetabarprime(\bfy)]_F &=& 
\tilde{Q}_-\left(1+ \bar{d}_s a{\bf D}_s \right) \bar{S}(0,\bfx;T,\bfy)
                                     \left(1-\bar{d}_s a\lvec{{\bf D}}_s\right)\tilde{Q}_+\,,\\[0ex]
[\zetaprime(\bfx) \zetabar(\bfy)]_F &=&  
\tilde{Q}_+\left(1+ \bar{d}_s a{\bf D}_s \right)  
                               \bar{S}(T,\bfx;0,\bfy)\left(1-\bar{d}_s a\lvec{{\bf D}}_s\right)\tilde{Q}_-\,,\\[0ex]
[\zetaprime(\bfx) \zetabarprime(\bfy)]_F &=& 
\tilde{Q}_+\left(1+ \bar{d}_s a{\bf D}_s \right) \Bigl[\bar{S}(T,\bfx;T,\bfy)\nonumber\\
    &&  - (\tilde{z}_f +\tilde{d}_s a{\bf D}_s)\delta(\bfx-\bfy)\Bigr]
                                     \left(1-\bar{d}_s a\lvec{{\bf D}}_s\right)\tilde{Q}_+\,,
\label{eq:contact2}
\end{eqnarray}
where  $\lvec{{\bf D}}_s = \frac12\left(\lvec{\nabla}_k+\lvec{\nabla}_k^\ast\right)\gamma_k$ 
is the counterpart to ${\bf D}_s$ in Eq.~(\ref{eq:bfDs}).
This defines the building blocks for any fermionic correlation function and therefore completes
the set-up for the massless theory. It also implicitly defines the fermionic boundary fields 
as integrands in the functional integral. For instance, one has
\begin{eqnarray}
   \zeta(\bfx)    &=& \left(1+ \bar{d}_s a{\bf D}_s\right) U_0(0,\bfx)\tilde{Q}_-\psi(a,\bfx),\\
   \zetabar(\bfx) &=& \psibar(a,\bfx)\tilde{Q}_- U_0(x)^\dagger\left(1- \bar{d}_s a \lvec{{\bf D}}_s\right), 
\end{eqnarray}
except that these equations do not take into account the contact terms proportional 
to $\tilde{z}_f$ and $\tilde{d}_s$, which need to be subtracted whenever the possible 
Wick contractions yield the  2-point functions $[\zeta(\bfx)\zetabar(\bfy)]_F$ 
or $[\zetaprime(\bfx)\zetabarprime(\bfy)]_F$ at coinciding spatial points 
$\bfx=\bfy$. 

The 2 renormalisation constants $z_f$ and $\tilde{z}_f$ and the 3 O($a$) 
improvement coefficients $d_s$, $\tilde{d}_s$  and $\bar{d}_s$ implement lattice variants 
of the 5 counterterms $\op{i}$ with $i=2,3,4,5,10$. The only missing counterterm is $\op{1}$, which can
can be taken care of by a multiplicative renormalization of the fermionic boundary fields,
\begin{equation}
   \zeta_{\rm R}=Z_\zeta\zeta, \qquad \zetabar_{\rm R}=Z_\zeta\zetabar, \qquad \zetaprime_{\rm R}=Z_\zeta\zetaprime,
   \qquad \zetabarprime_{\rm R}=Z_\zeta\zetabarprime,
\end{equation}
in complete analogy to the standard SF~\cite{Sint:1995rb,Luscher:1996sc}. 
While the O($a$) improvement coefficients $d_s$ and $\tilde{d}_s$ have their
standard SF counterparts in $\tilde{c}_t$ and $\tilde{c}_s$, the renormalization constants $z_f$ and $\tilde{z}_f$ and
the coefficient $\bar{d}_s$ are new and arise as a consequence of breaking the parity and flavour symmetries.
However, the situation is not as bad as it seems, since correlation functions involving 
$\tilde{z}_f$ and $\tilde{d}_s$ can be avoided in practice. Moreover, 
the counterterm proportional to $\bar{d}_s$ is $\gamma_5\tau^1$-odd and only contributes O($a^2$) effects
to $\gamma_5\tau^1$-even correlation functions, thus rendering it irrelevant for most applications.
In practice, this leaves us with the tuning of the renormalisation constant $z_f$ as the main new feature
as compared to the standard SF.

\subsection{Away from the chiral limit}

The possible counterterm structures in the presence of mass parameters can be obtained 
by applying the equations of motion for massive fermions 
to the counterterm basis obtained for the massless theory. Possible counterterms are
then given by those combinations which are invariant under charge conjugation.
Considering the mass terms 
\begin{equation}
   m\psibar\psi + \muq \psibar i\gamma_5\tau^3\psi,
\end{equation}
one finds that the boundary counterterms of dimension 4 have the form,
\begin{equation}
   \{m,\muq\}\times\left\{\op{1},\op{2},\op{3}\right\},
\end{equation}
and are thus proportional to the existing counterterms of dimension 3. 
They can be absorbed in a mass dependent rescaling of the corresponding
renormalisation constants $Z_\zeta$, $z_f$ and $\tilde{z}_f$.
It should be emphasised that a non-zero standard mass term entails the loss
of automatic O($a$) improvement. In order to maintain O($a$) improvement away from the
chiral limit one thus needs to introduce all the usual O($a$) bulk counterterms, 
just like in the standard SF. This is to be contrasted with a twisted mass parameter, 
which preserves the $\gamma_5\tau^1$-symmetry and is therefore compatible with automatic O($a$) improvement.
 
An interesting alternative mass term is obtained by introducing a twisted mass term in
a different flavour direction\footnote{The author thanks Jenifer Gonzalez Lopez for reminding him of this option.}, 
\begin{equation}
   \muq' \psibar i\gamma_5\tau^2\psi.
\end{equation}
Such a mass term is invariant under the chiral rotation~(\ref{eq:chiral_rot}) 
relating the standard and the chirally rotated SF, 
and is thus equivalent to a twisted mass parameter in the standard SF. 
Interestingly, in this case one needs a single countertem, which can be chosen to be 
\begin{equation}
   \muq'\left(\psibar\,\tilde{Q}_+ i\gamma_0\gamma_5\tau^2\psi - \psibar\,\tilde{Q}_-i\gamma_0\gamma_5\tau^2\psi\right).
\end{equation}
Note that this term is $\gamma_5\tau^1$-odd, so that it can be ignored, provided that 
$m=0$ and attention is restricted to $\gamma_5\tau^1$-even observables.
Thus, no additional O($a$) mass counterterm is needed whilst automatic
O($a$) improvement is maintained. 
The obvious drawback consists in the fact that the action ceases to be flavour diagonal.

%% file: section6.tex
\section{Renormalization and O($a$) improvement}

The basic 2-point functions in the free theory can be used to determine the
renormalization constants $z_f$ and $\tilde{z}_f$ 
and the O($a$) improvement coefficients $d_s$, $\tilde{d}_s$ and $\bar{d}_s$ 
at the tree level of perturbation theory. Some insight is provided for the r\^ole played by
the renormalisation constant $z_f$. Automatic O($a$) improvement is then
demonstrated with the free quark propagator in an Abelian background field.

\subsection{Interpretation of $z_f$ and $\tilde{z}_f$}

The renormalization constants $z_f$ and $\tilde{z}_f$ are related to the counterterms
$\op{2}$ and $\op{3}$ which are $\gamma_5\tau^1$-odd. When chirally rotated
back to the standard SF basis these 2 operators become proportional 
to $\psibar'P_\pm i\gamma_5\tau^3\psi'$, which indicates a breaking of parity and flavour symmetry. 
Indeed, in terms of the standard SF fields the $\gamma_5\tau^1$-transformation becomes 
a discrete SU(2) flavour transformation,
\begin{equation}
  \psi'\rightarrow -\tau^2 \psi'\qquad
  \psibar'\rightarrow -\psibar' \tau^2.
\end{equation}
Since parity and flavour symmetries are good continuum symmetries one
expects that both renormalization constants are finite, i.e.~scale independent functions of
the bare coupling, with an expansion of the form
\begin{equation}
 z(g_0) = z^{(0)} + g_0^2 z^{(1)} + \rmO(g_0^4).
\label{eq:coeffPT}
\end{equation}
It is straightforward to determine their lowest order values, by
comparing the basic 2-point functions with a vanishing gauge background field
to their expected continuum limit (cf.~Appendix~B). The free quark propagator on the
lattice is obtained analytically for $z_f=1$. The fact that 
this yields the expected continuum quark propagator up to cutoff effects 
immediately implies
\begin{equation}
  z_f^{(0)} = 1,
  \label{eq:zf0}
\end{equation}
as the correct tree-level value. Furthermore, the absence of contact terms in the continuum 
counterparts~(\ref{eq:2pt6}) and (\ref{eq:2pt9}) 
of the boundary-to-boundary 2-point functions~(\ref{eq:contact1},\ref{eq:contact2}) implies
\begin{equation}
\tilde{z}_f^{(0)}=\frac12.
\label{eq:tzf0}
\end{equation}
Note that these 2 finite renormalization constants restore the very 
$\gamma_5\tau^1$-symmetry which enables automatic O($a$) improvement. 
A possible strategy to determine $z_f$ and $\tilde{z}_f$ therefore consists in the requirement 
that 2 suitable $\gamma_5\tau^1$-odd observables vanish exactly.
Note that this strategy has implicitly been applied to obtain the results in Eqs.~(\ref{eq:zf0},\ref{eq:tzf0}) 
since the $\gamma_5\tau^1$-odd parts of the continuum 2-point functions all vanish identically. 
It should also be emphasized that $z_f$ affects any observable via the Wilson-Dirac operator,  
whereas $\tilde{z}_f$ only appears in the additive counterterm to the 2-point functions, Eqs.~(\ref{eq:contact1},\ref{eq:contact2}).
Hence the determination of $\tilde{z}_f$ can be avoided if the observables of interest do not involve 
these 2-point functions. This can often be arranged by choosing appropriate 
quark flavour assignments~\cite{LederSint}.

Another perspective on $z_f$ is obtained by considering an infinitesimal perturbation $\delta$ 
around the lowest order value, Eq.~(\ref{eq:zf0}). This would e.g.~be the situation
in perturbation theory with $g_0^2$ as the infinitesimal parameter.
The first order correction is equivalent to a single counterterm insertion in 
the correlation function of interest. Considering the quark propagator in the 
time-momentum representation and neglecting O($a$) artefacts, 
one obtains the expansion
\begin{eqnarray}
   \left.S^\tau_\bfp(x_0,y_0)\right\vert_{z_f=1+\delta} &=&  S_\bfp^{(\pi/2)}(x_0,y_0) - \delta 
   \Bigl[S_\bfp^{(\pi/2)}(x_0,0)S_\bfp^{(\pi/2)}(0,y_0) \nonumber \\ 
&& +\,S_\bfp^{(\pi/2)}(x_0,T)S_\bfp^{(\pi/2)}(T,y_0)\Bigr] 
+ \rmO\left(a,\delta^2\right),
\label{eq:proprot}
\end{eqnarray}
where the value of the angle $\alpha$ in the continuum propagator has been indicated by
a superscript. Interestingly, the first order correction can be expressed as a derivative of the
continuum propagator with respect to the angle $\alpha$,
\begin{equation}
 S_\bfp^{(\pi/2)}(x_0,0)S_\bfp^{(\pi/2)}(0,y_0)+S_\bfp^{(\pi/2)}(x_0,T)S_\bfp^{(\pi/2)}(T,y_0) 
= -2\frac{\partial}{\partial\alpha}S_\bfp^{(\alpha)}(x_0,y_0)\Bigl\vert_{\alpha=\pi/2}\Bigr..
\end{equation}
This gives rise to a complementary point of view regarding the r\^ole of $z_f$: 
starting from a free theory with boundary conditions involving the projectors $P_\pm(\alpha=\pi/2)=\tilde{Q}_\pm$,
the interactions tend to rotate the angle $\alpha$ away from $\pi/2$. 
Since there is no lattice symmetry which protects the angle $\alpha=\pi/2$,
one needs to tune $z_f$ to maintain the angle at $\pi/2$ all the way to the continuum limit.
This interpretation is further corroborated by a numerical experiment: 
one may tune $z_f$  such that the free lattice propagator $S^\tau_\bfp(x_0,y_0)$
approaches the continuum propagator defined at arbitrary values of $\alpha$, as long
as both time arguments are in the range $0<x_0,y_0<T$.

In conclusion, the finite renormalization constants $z_f$ and $\tilde{z}_f$ restore the 
$\gamma_5\tau^1$-symmetry, which is interpreted as a flavour symmetry in the standard SF basis. 
With the chosen set-up for the fermionic boundary fields the tuning of $z_f$ 
is equivalent to a renormalisation of the angle $\alpha$ such that the boundary 
conditions remain the ones with projectors $\tilde{Q}_\pm$.  In practice $z_f$  and $\tilde{z}_f$ can be fixed
by tuning suitable $\gamma_5\tau^1$-odd observables to zero. In analogy with
the O($a$) uncertainty in the definition of the chiral limit (cf.~Sect. 2), one expects
different determinations of $z_f$ or $\tilde{z}_f$ to differ by terms of O($a$). 
However, as in the case of the mass term, any O($a$) differences are expected to cause only 
O($a^2$) effects in $\gamma_5\tau^1$-even observables.

\subsection{O($a$) improvement coefficients at tree-level}

To obtain O($a$) improvement at the tree level one needs to adapt the counterterm coefficients such
that the continuum limit of the basic 2-point functions is approached with O($a^2$) corrections. 
All coefficients can be expanded in powers of $g_0^2$ in analogy to 
the finite renormalization constants, Eq.~(\ref{eq:coeffPT}). The first step should be 
the determination of $d_s$ since this counterterm enters the Wilson-Dirac operator
and thus affects the fermion propagator in the bulk. Note that the exact expressions for the fermion
propagator in Appendix~B have been calculated without the counterterm $\delta {\cal D}_W$ (\ref{eq:deltaDW}),
and therefore correspond to the choice $d_s=1$. Calculating the O($a$) effects in the
propagator for this case yields, in terms of the coefficient functions of Eqs.~(\ref{eq:Gpmlatt},\ref{eq:Hpmlatt}),
\begin{eqnarray}
G_\pm^\tau(\bfp;x_0,y_0) &=& G_\pm(\bfp;x_0,y_0) + \frac{\tau a}{2\cosh^2\omega T}\bigl\{\cosh\omega(x_0-y_0)\nonumber\\
&&\hphantom{0123456} \mp\,\sinh\omega T\sinh \omega(x_0+y_0-T)\bigr\} +\rmO(a^2),
\label{eq:GpmOa}\\
H_\pm^\tau(\bfp;x_0,y_0) &=& H_\pm(\bfp;x_0,y_0) + \frac{\tau a\omega}{2\cosh^2\omega T}\bigl\{\sinh\omega(x_0-y_0)\nonumber\\
&& \hphantom{0123456}\mp\,\sinh\omega T\cosh \omega(x_0+y_0-T)\bigr\} +\rmO(a^2).
\label{eq:HpmOa}
\end{eqnarray}
We thus see that for $\tau=0$ the O($a$) effects vanish so that $d_s^{(0)}=1$ is the correct choice in this
case. However, for $\tau=\pm 1$, there are uncancelled O($a$) effects, implying that
$d_s^{(0)}=1$ is incorrect in this case. In order to retain $d_s^{(0)}$ as a free parameter in the lattice 
propagator to O($a$), one may proceed by calculating a single insertion of this counterterm into the propagator, viz.
\begin{eqnarray}
  S^\tau_\bfp(x_0,y_0)\Bigl\vert_{d_s=d_s^{(0)}}\Bigr. &=& S^\tau_\bfp(x_0,y_0)\Bigl\vert_{d_s=1}\Bigr. \nonumber\\
&&-\, a\left(d_s^{(0)}-1\right)\Bigl[S_\bfp^{(\pi/2)}(x_0,0)i p_k^+\gamma_k S_\bfp^{(\pi/2)}(0,y_0)\nonumber\\ 
&&\hphantom{012}+\, S_\bfp^{(\pi/2)}(x_0,T)i p_k^+\gamma_k S_\bfp^{(\pi/2)}(T,y_0)\Bigr] +\rmO(a^2).
\end{eqnarray}
Hence the O($a$) effects in the first term of the r.h.s., explicitly given by Eqs.~(\ref{eq:GpmOa},\ref{eq:HpmOa})
should be cancelled by the second term, which may be calculated directly in the continuum. 
After some algebra one finds that in all cases 
\begin{equation}
   d_s^{(0)}=1-\frac{\tau}{2}, 
  \label{eq:ds0}
\end{equation}
appears to be the correct choice\footnote{Note that for this equation to hold with $\tau=-1$ it is assumed that the
counterterm $\delta{\cal D}_W$~(\ref{eq:deltaDW}) has been added to the Wilson-Dirac operator at 
Euclidean times $x_0=a,T-a$.}. This result has been confirmed by 
numerical inversion of the improved Wilson-Dirac operator 
in time-momentum space for a range of lattice resolutions, $L/a$. 

Having determined the O($a$) improved bulk propagator, it is straightforward to insert it into the
basic 2-point functions. Again we limit ourselves to $\tau=0$ and $\tau=+1$. It turns out that
$\bar{d}_s^{(0)}=0$ in both cases. Finally, requiring the contact terms in Eqs.~(\ref{eq:contact1},\ref{eq:contact2}) to be 
absent at O($a$) yields
\begin{equation}
  \tilde{d}_s^{(0)}=1.
\end{equation}

\subsection{A tree-level test of automatic O($a$) improvement}

In the free theory the simplest non-trivial correlation function is the fermion propagator in a fixed external 
gauge field.  We are thus led to consider ``observables" of the type
\begin{equation}
   I^{\pm}_{\Gamma_A} = -a^3\sum_{\bfx}\rme^{-i\bfp(\bfx-\bfy)} [\psibar(y)\Gamma_A\tilde{Q}_\pm\psi(x)]_F,
  \label{eq:tree-obs}
\end{equation}
where $\Gamma_A$ denotes any of the 16 linearly independent Hermitian $4\times 4$-matrices, 
$\gamma_\mu,1,\gamma_5,i\gamma_\mu\gamma_5,\sigma_{\mu\nu}$ (for $\mu,\nu=0,1,2,3$ and $\mu<\nu$).
Note that such bilocal quark-bilinear fields have a definite $\gamma_5\tau^1$-parity, which is
is even (odd) if $\Gamma_A$ anti-commutes (commutes) with $\gamma_5$.
Integrating over the fermions leads to
\begin{equation}
   I^{\pm}_{\Gamma_A}= \tr\left\{\Gamma_A\tilde{Q}_\pm S^{\tau=1}_{\bfp}(x_0,y_0)\right\},
\end{equation}
where the trace is over flavour, spin and colour indices, as these are  
contracted in Eq.~(\ref{eq:tree-obs}).
In the continuum limit, a chiral rotation relates these observables to standard SF correlators
(cf. Sect.~3). With vanishing mass parameters and in the continuum limit one thus expects
\begin{equation}
   I^{\pm}_{\Gamma_A,{\rm cont}} = \tr\left\{\Gamma_A P_\pm S^{(\alpha=0)}_{\bfp}(x_0,y_0)\right\},\qquad  
  \text{if $\{\gamma_5,\Gamma_A\}=0$}, 
\end{equation}
whereas $\gamma_5\tau^1$-odd observables vanish identically, due to flavour and parity symmetry 
(at a more technical level, in the standard SF basis, such observables are proportional to $\tr(\tau^3)=0$).
It is instructive to consider the same observables in the standard SF at finite lattice spacing, which 
will be denoted by ${I'}^{\pm}_{\Gamma_A}$.

To take the continuum limit we keep all dimensionful parameters ($x_0,y_0,T,\ldots$),
fixed in units of $L$. Furthermore, we assume that boundary O($a$) improvement is correctly implemented.
i.e.~that the O($a$) improvement coefficients, $d_s$ or, in the case of the standard SF,  $\tilde{c}_{\rm t}$, are set to their known
tree-level values $d_s^{(0)}=1-\tau/2$ (\ref{eq:ds0}) and $\tilde{c}^{(0)}_{\rm t}=1$~\cite{Luscher:1996sc}. 
For $\Gamma_A$'s which anti-commute with $\gamma_5$ one then expects
\begin{eqnarray}
  \left.{I'}^{\pm}_{\Gamma_A}\right\vert_{\{\Gamma_A,\gamma_5\}=0} 
   &=& I^{\pm}_{\Gamma_A,{\rm cont}} + const \times(\csw-1)\frac{a}{L} +\rmO(a^2),
\label{eq:stdSF_e}\\
   \left.I^{\pm}_{\Gamma_A}\right\vert_{\{\Gamma_A,\gamma_5\}=0} 
   &=& I^{\pm}_{\Gamma_A,{\rm cont}} +\rmO(a^2).
\label{eq:chiSF_e}
\end{eqnarray}
While the standard SF observables in Eq.~(\ref{eq:stdSF_e}) require $\csw=1$ to be O($a$) improved,
$\chi$SF observables in Eq.~(\ref{eq:chiSF_e}) should reach the same continuum limit with 
a rate of $\rmO(a^2)$, regardless of the value of $\csw$. 
On the other hand, for $\gamma_5\tau^1$-odd observables one expects,
\begin{eqnarray}
  \left.{I'}^{\pm}_{\Gamma_A}\right\vert_{[\Gamma_A,\gamma_5]=0} 
   &=& 0,
\label{eq:stdSF_o}\\
   \left.I^{\pm}_{\Gamma_A}\right\vert_{[\Gamma_A,\gamma_5]=0} 
   &=&  const \times(\csw-1)\frac{a}{L} + \rmO(a^2).
\label{eq:chiSF_o}  
\end{eqnarray}
Like in the continuum, the exact flavour symmetry in the standard
SF implies that $\gamma_5\tau^1$-odd observables are exactly zero [Eq.~(\ref{eq:stdSF_o})].
While some of the $\chi$SF observables vanish exactly, too,
most are afflicted by O($a$) effects which can be reduced to O($a^2$) by Symanzik 
improvement with $\csw=1$, as indicated in Eq.~(\ref{eq:chiSF_o}).

In order to see the effect of the Sheikholeslami-Wohlert term at tree-level one needs 
the presence of a background gauge field. We will use the same
constant Abelian SU(2) background gauge field which has been used for the definition of the SF coupling~
in the SU(2) Yang-Mills theory \cite{Luscher:1992zx}. 
The fermion propagator in the continuum is known analytically in this case; however its lattice
counterparts have so far only been calculated in the standard SF~(cf.~ref.~\cite{Luscher:1996vw}).
We therefore resort to a numerical approach. We choose the set-up with $\tau=+1$, 
set $T=L$ and $\bfp ={\bf 0}$, but keep a non-zero value $\theta=1.0$. 
Specifying the Euclidean times $x_0=T/2$ and $y_0=T/4$ 
we choose the $+$-components with $\Gamma_A=\gamma_0$ and $\Gamma_A=1$ 
as examples of $\gamma_5\tau^1$-even and -odd quantities, respectively.
Using the analytic result for the continuum propagator of Appendix~B, their continuum limit
can be calculated analytically, with the result,
\begin{equation}
  I_{\gamma_0,{\rm cont}}^+ = 4 \sum_{i=1}^2 \frac{\cosh\left[\omega_i(T/2)-\omega_i(T)\right]
\cosh \omega_i(T/4)}{\cosh\omega_i(T)},
\end{equation}
whereas $I_{1,{\rm cont}}^+ = 0$. The sum over colour components corresponds to the colour trace in the
special case of the SU(2) Abelian background field. Computing these observables  
for $\csw=0$ and $\csw=1$, and lattice sizes from $L/a=8$ up to $L/a=72$,
universality is indeed confirmed. In Figure~1 the relative deviations from the
common continuum limit, rescaled by $L/a$, are plotted versus $a/L$. This extrapolates to
a finite value in the case of O($a$) corrections, whereas a vanishing continuum limit  
indicates O($a^2$) or higher lattice artefacts. 
\begin{figure}[h]
 \epsfig{file=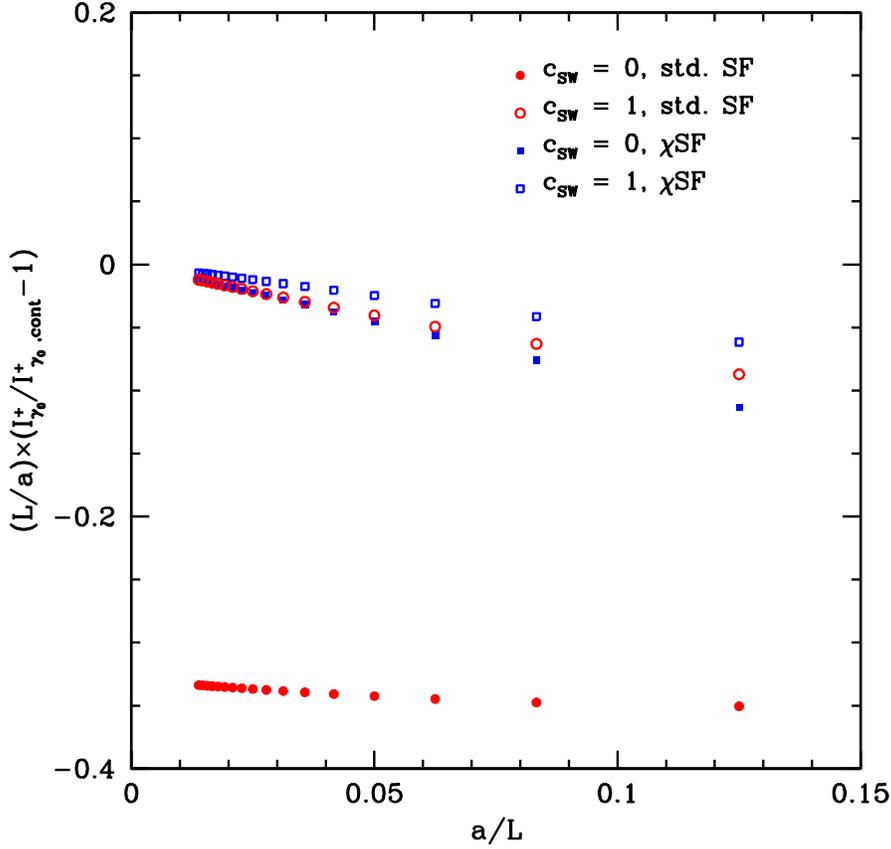,width=12cm}
 \caption{The relative deviation from the continuum limit, rescaled by $L/a$, is plotted vs.~$a/L$ 
for the $\gamma_5\tau^1$-even observable $I^+_{\gamma_0}$ (cf.~text for further details).}
\end{figure}
Finally, Fig.~2 shows the $\gamma_5\tau^1$-odd observable $I^+_1$, multiplied by $L/a$.
Hence, a finite value in the continuum limit indicates O($a$) effects. This is the generic situation
unless $\csw=1$ for which the observable becomes a pure lattice artefact of O($a^2$). 
In the free theory we have thus explicitly verified that automatic O($a$) improvement works as expected.
\begin{figure}[h]
 \epsfig{file=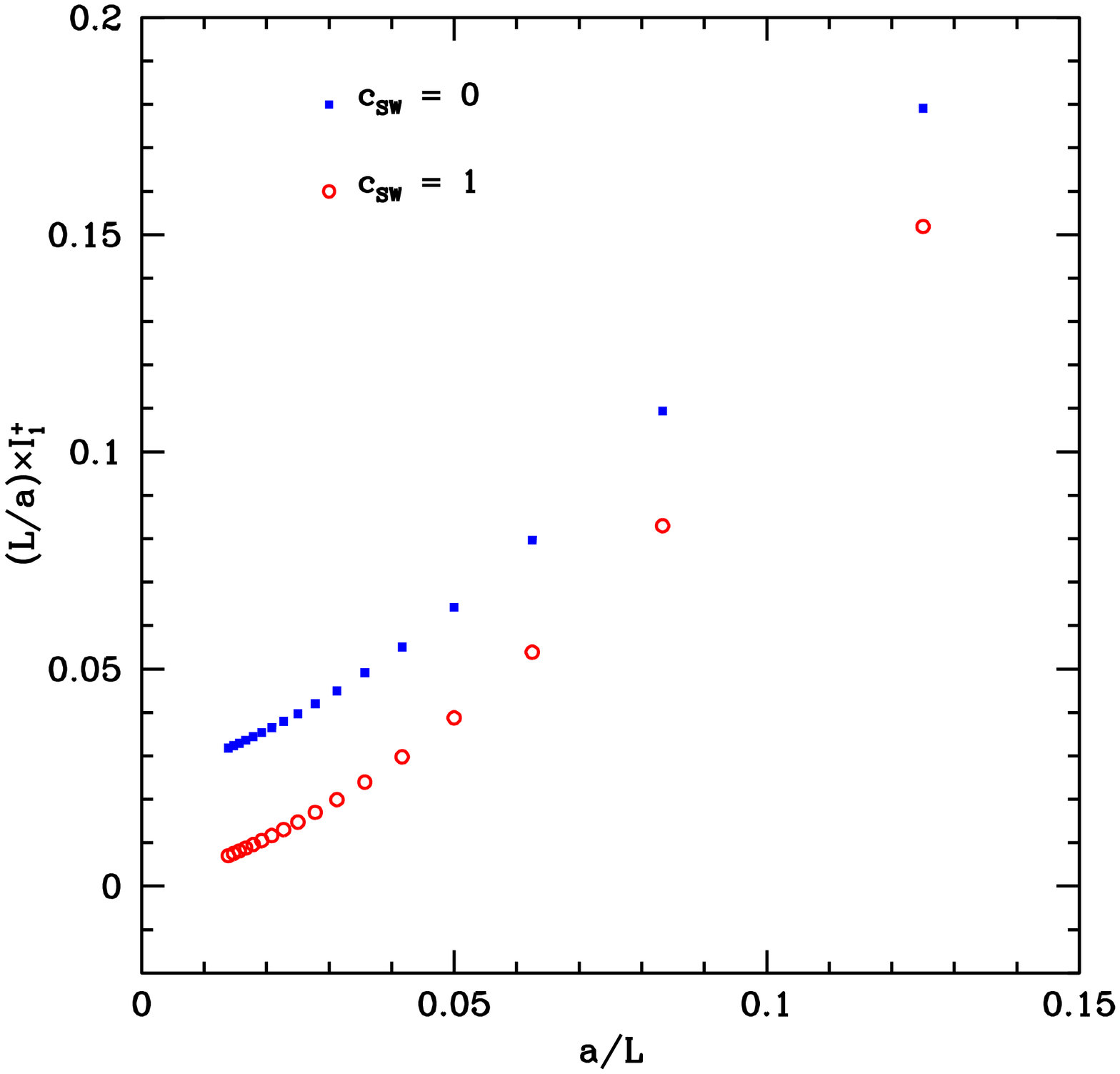,width=12cm}
 \caption{The O($a$) effect in the $\gamma_5\tau^1$-odd observable $I_1^+$ with $\csw=0$ is rendered visible 
by rescaling the observable with $L/a$. The effect of O($a$) improvement with $\csw=1$ reduces the
lattice artefacts to O($a^2$).}
\end{figure}

%% file: section7.tex
\section{Concluding remarks and outlook}

We have shown how the mechanism of automatic O($a$) improvement 
can be reconciled with SF boundary conditions. This means that any bulk O($a$) effects vanish
in correlation functions which are even under the $\gamma_5\tau^1$-symmetry. Any remaining
O($a$) effects must arise from boundary effects and their elimination requires  the 
tuning of a couple of boundary counterterms (proportional to $d_s$ and $c_t$), similar to
the standard SF. In order to achieve this one needs to tune the coefficient $z_f$ of a 
dimension 3 boundary counterterm such that the physical parity symmetry is restored.
One may say that the tuning problem for $z_f$  is the price to be paid
for avoiding the non-perturbative determination of $\csw$ and of any O($a$) improvement
coefficients for composite fields in the correlation functions 
(such as $\ca$ or $\cv$ for the flavour currents~\cite{Jansen:1995ck}).
The chirally rotated SF  provides a well-defined framework which can be seen as
an alternative regularisation of the standard SF, with the advantage 
of reduced cutoff effects. This will benefit the calculation of step-scaling functions
for which the continuum extrapolation will be made easier, and it will reduce the potential
problem of introducing O($a$) effects into hadronic matrix elements through SF scheme 
renormalization factors.

Since the formulation of the chirally rotated SF was given for a flavour doublet, 
a natural question to ask is whether it can be generalised to arbitrary flavour number $\Nf$.

\subsection{General $\Nf$}

Most of the discussion in this paper concerned the case of a quark flavour doublet, where
a non-singlet axial rotation relates the correlation functions with those 
defined in the standard SF. In particular, invoking universality, one expects 
that the renormalisation works out in the same way as in the standard SF, 
up to scale-independent (finite) counterterms, which are the consequence of 
breaking the continuum flavour and parity symmetry by the regularisation.
This is further supported by the formulation with Ginsparg-Wilson quarks, where 
the same symmetries are restored \`a la Ginsparg-Wilson~\cite{Sint:2007zz}.

While the inclusion of further doublets and thus the formulation for 
even flavour numbers is straightforward, the situation is a bit more complicated 
for odd $\Nf$. First of all, the discussion of
the symmetries, and hence the counterterm structure, relied on a flavour exchange between 
up- and down-type flavours combined with space-time reflections.
To apply this analysis to the single flavour case one may use a spurion field in the
Wilson-Dirac operator, which switches sign under a space or time reflection, 
and thus takes over the r\^ole of the flavour exchange symmetry. 
The counterterm structure found in this way would then be the same as in the doublet case,
appropriately reduced to either the up- or down-type flavour. This may in fact be generalised
to any number $\Nf$ of flavours, with all flavours being of the up-flavour type, for example.
As the relation to the standard SF is lost in this case, one would then tend to 
describe the symmetries as an exact SU($\Nf$) flavour symmetry, 
with all axial symmetries being broken by a singlet twisted mass term with O(1) coefficient, 
localised at the boundaries. In addition to the axial symmetries also parity and time reflection 
symmetries are broken explicitly by the boundary conditions. Note, however, that the interpretation 
that $z_f$ restores a continuum symmetry is lost in this case, and it is unclear to the author
whether $z_f$ remains scale-independent in this situation. A further, 
more practical question is whether such a set-up would be suitable for numerical simulations.

\subsection{Numerical simulations including the fermion determinant}

Currently, most numerical simulations are based on 
some variant of the Hybrid Monte Carlo algorithm~\cite{Duane:1987de}, and the fermionic determinant 
is thus required to be real and positive.
For the Wilson-Dirac operator in the two-flavour case with $\chi$SF boundary conditions 
this property can easily be established. The hermiticity of $\gamma_5\tau^1 {\cal D}_W$, 
implies for its flavour components,
\begin{equation}
  {\cal D}_W = \diag\left({\cal D}_W^{(1)},{\cal D}_W^{(2)}\right), \qquad 
   \gamma_5 {\cal D}_W^{(1)}\gamma_5=\left({\cal D}_W^{(2)}\right)^\dagger,
\end{equation}
and therefore,
\begin{eqnarray}
   \det \left( {\cal D}_W+m_0 \right) &=& \det\left({\cal D}_W^{(1)}+m_0\right) 
   \det \left( {\cal D}_W^{(2)}+m_0\right) \nonumber\\ 
&=&  \det \left[\left({\cal D}_W^{(1)}+m_0\right)^\dagger \left({\cal D}_W^{(1)}+m_0\right)\right] \geq 0.
\label{eq:det}
\end{eqnarray}
which has  exactly the structure required for the standard algorithms.
Twisted mass terms can be added, too. Allowing for two kinds of twisted mass terms, 
\begin{equation}
i\muq\gamma_5\tau^3 + i\muq'\gamma_5\tau^2,
\end{equation}
one realizes that the Wilson-Dirac operator is still $\gamma_5\tau^1$-hermitian 
so that its determinant must be real. Reducing the determinant in flavour space 
the r.h.s.~of Eq.~(\ref{eq:det}) generalises as follows, 
\begin{equation}
 \det \left[\left({\cal D}_W^{(1)}+m_0\right)^\dagger \left({\cal D}_W^{(1)}+m_0\right)
  +i\muq\gamma_5\left({\cal D}_W^{(1)}-{\cal D}_W^{(2)}\right) +\muq^2+ (\muq')^2\right].
\end{equation}
Note that $\muq$ multiplies a hermitian operator localised at the boundaries, which is 
diagonal in space-time and gauge field independent, at least  
for the set-up with $\tau=\pm 1$ (cf.~Sect.~4).

The situation is less clear with odd numbers of flavours. In the free theory the determinant
for a single flavour turns out to be real, even in the presence of a spatially constant Abelian 
background field. However, a numerical check on a small lattice with a random gauge configuration
reveals that the determinant is in general complex\footnote{I thank B.~Leder for performing this
numerical check using MATLAB.}. Hence, some additional idea is needed in order to enable simulations of  
odd flavour numbers with $\chi$SF-type boundary conditions.
A simple alternative in this case would be a mixed formulation, where an even number of quarks
is represented by chirally rotated quark doublets, whereas a single quark is implemented with
standard SF boundary conditions. In the continuum limit this formulation will be equivalent to
the standard SF with an odd number of flavours. Moreover, automatic O($a$) improvement 
would  partially apply: while the Sheikholeslami-Wohlert term in the action would be required,
correlation functions involving only chirally rotated quark flavours would be automatically 
O($a$) improved up to boundary effects. The price to pay is an additional O($a$) improvement 
coefficient at the boundaries since now both $d_s$ and $\tilde{c}_{\rm t}$ would be required.

\subsection{Outlook} 

The chirally rotated SF offers a number of applications. Automatic O($a$) improvement 
will be particularly welcome when applying SF schemes e.g.~to four-quark operators  
for which a non-perturbative determination of the O($a$) counterterms seems impractical. 
Furthermore, in the standard SF, the O($a$) improvement by the Sheikholeslami-Wohlert 
term often cancels large cutoff effects.  In order to apply SF schemes 
to operators used in twisted mass QCD without the Sheikholeslami-Wohlert term 
(as currently used by the European twisted mass collaboration~\cite{Baron:2010bv}),
the chirally rotated SF is a natural choice, and may indeed be the only practical one.
Further applications include multi-flavour QCD-like theories or 
lattice gauge theories with different gauge groups and fermion representations,
where the determination of the Sheikholeslami-Wohlert coefficient $\csw$ can be avoided. Moreover,
one may expect a reduced sensitivity to the precise definition of the critical quark mass.
Finally, the chirally rotated SF  provides new methods to determine finite renormalisation constants
such as the axial current normalisation constant $Z_{\rm A}$, and some of the usual O($a$) improvement 
coefficients~\cite{LederSint}. Since standard Symanzik O($a$) improvement of Wilson fermions 
remains relevant for many collaborations, any improvement over the traditional 
methods based on Ward identities may indeed be welcome.

However, before these applications can be envisaged, one would like to perform a thorough 
non-perturbative test of the chirally rotated SF. In particular one needs 
to acquire some experience regarding the tuning of $z_f$ and $\mcr$~\cite{Lopez:2009yc,LederSint}. 
One would like to verify that the boundary conditions are correctly implemented 
in the interacting theory and that automatic O($a$) improvement works out as anticipated.
Furthermore, universality could be checked by comparing with correlation
functions defined in the standard SF. A non-perturbative study
addressing all these questions is currently being performed in quenched lattice QCD~\cite{LederSint}.
Further checks will be carried out in perturbation theory~\cite{JaminSint}. In
particular, one would like to obtain perturbative estimates for $d_s$ and $c_{\rm t}$,
as these improvement coefficients are difficult to determine otherwise.

%% file: acknow.tex
\subsection*{Acknowledgments}
The author would like to thank Bj\"orn Leder for a pleasant collaboration on 
the numerical study of the chirally rotated SF, and Jenifer Gonzalez Lopez for useful feedback. 
Helpful discussions with Rainer Sommer and Martin L\"uscher are gratefully acknowledged. 
This article was finished during the CERN Theory Institute ``Future directions in lattice gauge theory". 
I am grateful to CERN and the organizers for financial support and the inspiring 
atmosphere at this workshop. This work was partially supported by the EU under Grant Agreement number 
PITN-GA-2009-238353 (ITN STRONGnet).

%% file: appendixA.tex
\appendix
\renewcommand{\thesection}{A  Absence of zero modes}
\section{}
\renewcommand{\thesection}{A}

The absence of zero modes for the massless continuum Dirac operator in a 
smooth external gauge field $A_\mu(x)$, can be established as in the case of the SF 
with standard boundary conditions~\cite{Luscher:2006df}. We first deal with the
case of a single quark flavour with $Q_\pm$ boundary conditions, 
and then consider the case of a quark doublet at generic values of the angle $\alpha$.

\subsection{Single flavour with $Q_\pm$ boundary conditions}

Consider the massless Dirac operator $\Dslash = \gamma_\mu(D_\mu+A_\mu)$ 
for a single flavour, acting in the Hilbert space of square integrable functions 
$\varphi(x)$ on a hyper cylinder, subject to the boundary conditions
\begin{equation}
    Q_+ \varphi(0,\bfx)=0 = Q_-\varphi(T,\bfx),\qquad  Q_\pm=\frac12(1\pm i\gamma_0\gamma_5).
\end{equation}
With respect to the natural scalar product,
\begin{equation}
  (\varphi,\chi) = \int_0^T\rmd x_0\int \rmd^3{\bf x}\, \varphi(x)^\dagger \chi(x),
\end{equation}
the operator $i\Dslash$ is hermitian. A well-defined eigenvalue problem is thus obtained,
\begin{equation}
  i\Dslash w_n = \lambda_n w_n,
\end{equation}
with real eigenvalues $\lambda_n$ and smooth eigenfunctions $w_n$, which span the Hilbert space.
In the following it is assumed that the eigenfunctions have been chosen such as to 
form an orthonormal basis of the Hilbert space, i.e.
\begin{equation}
  (w_n,w_m)=\delta_{nm},\qquad  \sum_n w_n(x) w_n(y)^\dagger = \delta^{(4)}(x-y).
\end{equation}
Next, consider the classical solution of the Dirac equation with inhomogeneous 
boundary conditions,
\begin{equation}
\Dslash\,\psi_{\rm cl} (x) = 0, \qquad  Q_+\psi_{\rm cl}(0,\bfx) = \rho(\bfx),\qquad 
Q_-\psi_{\rm cl}(T,\bfx) = \rho'(\bfx).
\end{equation}
In order to obtain an explicit expression for the classical solution,
one needs the propagator in the orthogonal subspace of the zero modes,
\begin{equation}
    S^\perp(x,y) = -i \sum_{n:\lambda_n\ne 0}\frac{w_n(x) w_n(y)^\dagger}{\lambda_n}.
\end{equation}
Assume that there are a finite number of $N_0$ zero modes, $w_{n_i}(x)$, $i=1,\ldots,N_0$. 
The propagator in the orthogonal subspace then satisfies the equation
\begin{equation}
 \Dslash S^\perp(x,y) = \delta^{(4)}(x-y) - \sum_{i=1}^{N_0} w_{n_i}(x) w_{n_i}(y)^\dagger.
\end{equation}
Furthermore, since any finite superposition of zero modes must be smooth, 
the discontinuity for $S^\perp(x,y)$ is given by
\begin{equation}
   S^\perp_>(x_0,\bfx,x_0,\bfy)-S^\perp_<(x_0,\bfx,x_0,\bfy) = \gamma_0\delta^{(3)}(\bfx-\bfy),
   \label{eq:disc}
\end{equation}
where the subscripts refer to the ordering of the time arguments in the propagator.
Using these properties, the classical solution can be written as follows,
\begin{equation}
   \psi_{\rm cl}(x) = \int\rmd^3\bfy \left[ S^\perp(x,0,\bfy)\gamma_0 Q_+\rho(\bfy) -
                                        S^\perp(x,T,\bfy)\gamma_0 Q_-\rho'(\bfy)\right],
\label{eq:classical}
\end{equation}
showing that it is linearly related to the boundary values.
Finally, note that any of the zero modes, $w_{n_i}(x)$, is a classical solution of the Dirac
equation, and can  be written in the form (\ref{eq:classical}). 
Since its boundary values vanish, the zero modes must vanish, too, i.e. $w_{n_i}=0$, $i=1,\ldots,N_0$. 

The case of opposite boundary projectors is completely analogous. Hence, the absence
of zero modes is guaranteed with any number of quark flavours, and for any mixture 
of up and down type flavours. 

\subsection{Flavour doublet at generic values of $\alpha$}

Essentially the same argument works for a flavour doublet at any value of $\alpha$,
except that the mathematical structure is a bit more complicated. 
The boundary conditions for $\psi$ and $\psibar$ at fixed, but arbitrary $\alpha$
give rise to the definition of two different Hilbert spaces,
\begin{equation}
   {\cal H}_\pm = \{ \phi: P_\pm(\pm\alpha)\phi\vert_{x_0=0}=0 = P_\mp(\pm\alpha)\phi\vert_{x_0=T}\},
\end{equation}
and both $\Dslash$ and $\gamma_5\tau^1$ can be seen as mappings between these two spaces.
Hence, $\gamma_5\tau^1\Dslash$ is a hermitian operator in either space, with a complete set
of smooth eigenfunctions and real eigenvalues. 
Denoting its orthonormal eigenfunctions in ${\cal H}_+$ by $\varphi_n$, 
\begin{equation}
 \gamma_5\tau^1\Dslash \varphi_n = \lambda_n \varphi_n,
\end{equation}
it follows that the functions $\tilde{\varphi}_n = \gamma_5\tau^1 \varphi_n$ 
form an orthonormal basis in ${\cal H}_-$, and
\begin{equation}
 \Dslash\gamma_5\tau^1 \tilde{\varphi}_n = 
\Dslash \varphi_n = \lambda_n\gamma_5\tau^1 \varphi_n= \lambda_n\tilde{\varphi}_n.
\end{equation}
Completeness in either Hilbert space then implies
\begin{equation}
   \delta^4(x-y) = \sum_n \varphi_n(x)\varphi_n(y)^\dagger 
   = \sum_n \tilde{\varphi}_n(x)\tilde{\varphi}_n(y)^\dagger.
\end{equation}
To relate the classical solution of the Dirac equation, $\Dslash\,\psi_{\rm cl} (x) = 0$,
to its boundary values,
\begin{equation}
P_+(\alpha)\psi_{\rm cl}(0,\bfx) = \rho(\bfx),\qquad 
P_-(\alpha)\psi_{\rm cl}(T,\bfx) = \rho'(\bfx),
\end{equation}
one needs again the inverse of $\Dslash$ on the non-zero modes,
\begin{equation}
    S^\perp(x,y) =  \sum_{n:\lambda_n\ne 0}\frac{\varphi_n(x) \tilde{\varphi}_n(y)^\dagger}{\lambda_n}.
\end{equation}
Assuming $N_0$ zero modes, this propagator satisfies
\begin{equation}
 \Dslash S^\perp(x,y) = \delta^{(4)}(x-y) 
- \sum_{i=1}^{N_0} \tilde{\varphi}_{n_i}(x)\tilde{\varphi}_{n_i}(y)^\dagger.
\end{equation}
The discontinuity is again described by Eq.~(\ref{eq:disc}), and the classical
solution takes the form
\begin{equation}
   \psi_{\rm cl}(x) = \int\rmd^3\bfy \left[ S^\perp(x,0,\bfy)\gamma_0 P_+(\alpha)\rho(\bfy) -
                                        S^\perp(x,T,\bfy)\gamma_0 P_-(\alpha)\rho'(\bfy)\right].
\label{eq:classical1}
\end{equation}
Again, any of the zero modes solves the Dirac equation and satisfies homogeneous boundary conditions,
and must therefore vanish identically. Note that this argument also holds at $\alpha=0$.
In particular, the flavour permutation $\tau^1$ is not required in this case, so that 
the proof applies virtually unchanged to the case of a single quark flavour 
with standard SF boundary conditions.

%% file: appendixB.tex
\appendix
\renewcommand{\thesection}{B  The free quark propagator in the continuum limit}
\renewcommand{\thesection}{B}
\section{The free quark propagator}
We collect a few formulae for the free propagator both in the continuum and on the lattice.
For the sake of notational simplicity the propagator is generically denoted by $S$ and
any additional parameter dependence or indeed whether it is the lattice or continuum propagator
will be either clear from the context or otherwise indicated.

\subsection{Continuum quark propagator for generic values of $\alpha$}
The free quark propagator in the continuum theory can be easily obtained using standard methods.  
Allowing for standard and twisted mass terms, the quark propagator satisfies,
\begin{equation}
   \left(\Dslash + m+i\muq\gamma_5\tau^3\right) S(x,y) = \delta^{(4)}(x-y),
\end{equation}
and the boundary conditions,
\begin{xalignat}{2}
     P_+(\alpha)S(x,y)\vert_{x_0=0} &=0,    
    &P_-(\alpha)S(x,y)\vert_{x_0=T} &=0,\nonumber\\
    S(x,y)P_-(-\alpha)\vert_{y_0=0} &= 0,   
    & S(x,y)P_+(-\alpha)\vert_{y_0=T} &= 0.
\end{xalignat}
The boundary conditions in the second argument follow e.g.~from the fact
that the Dirac operator is $\gamma_5\tau^1$-hermitian (cf.~Appendix~A).
For the same reason the complementary components satisfy (modified) Neumann
conditions, i.e.
\begin{xalignat}{2}
     (D_0-M_\alpha) P_-(\alpha)S(x,y)\vert_{x_0=0} &=0,    
    &(D_0+M_\alpha)P_+(\alpha)S(x,y)\vert_{x_0=T} &=0,\nonumber\\
    S(x,y)P_+(-\alpha)(\lvec{D}_0-M_\alpha)\vert_{y_0=0} &= 0,   
    & S(x,y)P_-(-\alpha)(\lvec{D}_0+M_\alpha)\vert_{y_0=T} &= 0,
\label{eq:rotbc_complementary}
\end{xalignat}
with
\begin{equation}
   M_\alpha   = m\cos\alpha+\muq\sin\alpha.
\end{equation}
These equations hold for arbitrary smooth gauge field $A_\mu(x)$. 
Setting the gauge field to zero, we use $L$-periodic spatial boundary conditions
and include the phase $\theta$ as an Abelian background field in the spatial derivative:
\begin{equation}
  \Dslash = \partial_0\gamma_0 + \left(\partial_k+i\frac{\theta}{L}\right)\gamma_k.
\end{equation}
Spatial translation invariance suggests to pass to the time momentum representation,
\begin{equation}
   S(x,y)= L^{-3}\sum_{\bfp}\rme^{i\bfp(\bfx-\bfy)} S_\bfp(x_0,y_0),
\end{equation}
where the (infinite) sum is over all allowed momenta of the form $\bfp=2\pi\bfn/L$,
with $\bfn$ being a triple of integers.
In time-momentum representation one then has
\begin{equation}
  \left(\partial_0 -A\right) S_\bfp(x_0,y_0) = \gamma_0\delta(x_0-y_0),
\end{equation}
with
\begin{equation}
   A = \left(i p_k^+\gamma_k-m+i\muq\gamma_5\tau^3\right)\gamma_0 = A^\dagger, \qquad  p_k^+=p_k+\theta/L.
\end{equation}
There are many ways to represent the free propagator. A very compact form is
\begin{equation}
S_\bfp(x_0,y_0)=\rme^{x_0A}\Bigl\{\frac{-1}{n(T)}P_-(\alpha)\rme^{TA}
+\theta(x_0-y_0)\Bigr\}\rme^{-y_0A}\gamma_0,
\end{equation}
where $\theta(t)$ denotes the Heaviside step function, and 
\begin{equation}
 n(T)=\cosh(\omega T)+ \frac{M_\alpha}{\omega}\sinh(\omega T),\qquad
 \omega = \sqrt{(\bfp^+)^2+m^2+\muq^2}.
\end{equation}
Using $A^2=\omega^2$ one may easily evaluate the exponentials and obtain a more 
explicit representation,
\begin{eqnarray}
   S_\bfp(x_0,y_0) &=& \gamma_0P_-(-\alpha)H_+ + \gamma_0P_+(-\alpha)H_-\nonumber\\
                   &&  -(A-M_\alpha)\gamma_0P_-(-\alpha)G_+ -(A+M_\alpha)\gamma_0P_+(-\alpha)G_-,
\end{eqnarray}
with coefficient functions $G_\pm$ and $H_\pm$ defined by
\begin{eqnarray}
 G_+(\bfp;x_0,y_0)&=&\frac{1}{2\omega R_\alpha}\Bigl\{(M_\alpha-\omega)\left(\rme^{\omega(|x_0-y_0|-T)}-\rme^{\omega(x_0+y_0-T)}\right)\nonumber\\
&&+\,(M_\alpha+\omega)\left(\rme^{-\omega(|x_0-y_0|-T)}-\rme^{-\omega(x_0+y_0-T)}\right)\Bigr\},\\
G_-(\bfp;x_0,y_0)&=& G_+(\bfp;T-x_0,T-y_0),\\
H_\pm(\bfp;x_0,y_0) &=& (-\partial_0\mp M_\alpha)G_\pm(\bfp;x_0,y_0),
\end{eqnarray}
and
\begin{equation}
   R_\alpha= (M_\alpha+\omega)\rme^{\omega T} -(M_\alpha-\omega)\rme^{-\omega T}.  
\end{equation}
Note that the coefficient functions are indexed such as to conform with the projectors for $\alpha=\pi/2$, where
$P_\pm(-\pi/2)=\tilde{Q}_\mp$. They enjoy the following properties:
\begin{equation}
   G_\pm(\bfp;x_0,y_0)=G_\pm(\bfp;y_0,x_0),\qquad H_\pm(\bfp;x_0,y_0)=-H_\mp(\bfp;y_0,x_0).
\end{equation}
The Dirichlet boundary conditions imply
\begin{equation}
   G_+(\bfp;0,y_0) = 0 = G_-(\bfp;T,y_0),
\end{equation}
and
\begin{equation}
   H_-(\bfp;0,y_0)= 0 = H_+(\bfp;y_0,0),\qquad  H_+(\bfp;T,y_0)= 0= H_-(\bfp;y_0,T).
\end{equation}
Of particular interest is the massless case, for which the 
formulae simplify,
\begin{eqnarray}
 G_\pm(\bfp;x_0,y_0) &=& \frac{1}{2\omega\cosh\omega T}
      \Bigl\{-\sinh\omega(|x_0-y_0|-T) \nonumber\\
       &&\hphantom{012345467890123456}\pm \sinh\omega(x_0+y_0-T)\Bigr\},
\label{eq:Gpmcont}\\
 H_\pm(\bfp;x_0,y_0) &=& \frac{1}{2 \cosh\omega T}\Bigl\{\varepsilon(x_0-y_0)\cosh\omega(|x_0-y_0|-T) \nonumber\\
 && \hphantom{0123454678901234}\mp \cosh\omega(x_0+y_0-T)\Bigr\},
\label{eq:Hpmcont}
\end{eqnarray}
where $\varepsilon(t)$ denotes the sign function.

\subsection{The massless continuum quark propagator in an Abelian and spatially constant background field}

In this subsection, an explicit expression is given for the massless continuum
quark propagator in an Abelian SU($N$) background field $B_\mu$. The latter is taken
to be spatially constant, i.e. $B_\mu=B_\mu(x_0)$, and in the $B_0=0$ gauge.
Spatial translation invariance then allows to pass to the time-momentum representation, 
where the propagator satisfies,
\begin{equation}
  \left(\partial_0\gamma_0+\left[ip_k^++B_k(x_0)\right]\gamma_k\right)S_\bfp(x_0,y_0)= \delta(x_0-y_0).
\end{equation}
We are interested in the background fields used for the running coupling~\cite{Luscher:1992an,Luscher:1992zx,Luscher:1993gh} or 
the determination of the Sheikholeslami-Wohlert coefficient~\cite{Luscher:1996ug}.
These are of the form
\begin{equation}
  B_k(x_0)=  C_k+\frac{x_0}{T} \left(C_k'-C_k\right),
\end{equation}
where the boundary gauge fields are taken to be
\begin{equation}
   C_k=i\phi/L,\qquad C_k'=i\phi'/L,
\end{equation}
with diagonal $N\times N$-matrices $\phi$ and $\phi'$ in colour space. Since these are chosen
to be independent of the spatial index $k$, it is convenient to restrict attention to the 
special case where also the momentum components are $k$-independent, i.e.~$p_k^+=p^+$ for $k=1,2,3$.
Then, defining
\begin{eqnarray}
  \Sigma &=& \frac{i}{\sqrt{3}}\left(\gamma_1+\gamma_2+\gamma_3\right)\gamma_0, \\
   \omega(t) &=& t \sqrt{3}\left(p^+ +\frac{\phi}{L} +\frac{t}{2LT}\left[\phi'-\phi\right]\right),
\end{eqnarray}
the result for the quark propagator can be written in the form
\begin{equation}
  S_{\bfp}(x_0,y_0) = \rme^{\omega(x_0)\Sigma}\left\{-\frac{1}{\cosh\omega(T)}P_-(\alpha)\rme^{\omega(T)\Sigma}
  + \theta(x_0-y_0)\right\}\rme^{-\omega(y_0)\Sigma}\gamma_0.
\end{equation}
where $\theta(t)$ denotes the Heaviside step function. Using $\Sigma^2=1$, one may easily
obtain the equivalent expression 
\begin{eqnarray}
  S_{\bfp}(x_0,y_0) &=& \gamma_0 P_-(-\alpha) H_+ +\gamma_0 P_+(-\alpha)_- H_- \nonumber\\
                    && \mbox{}-\omega'(T)\Sigma\gamma_0\left\{P_-(-\alpha) G_+ + P_+(-\alpha) G_-\right\}.
\end{eqnarray}
Here, $\omega'(t)\equiv\rmd\omega(t)/\rmd t$, and the coefficient functions are defined by
\begin{eqnarray}
G_\pm(\bfp;x_0,y_0) &=& \frac{1}{2\omega'(T)\cosh\omega(T)} \Bigl\{ \pm\sinh\left[\omega(x_0)+\omega(y_0)-\omega(T)
\right] \nonumber\\          && -\sinh\left[\varepsilon(x_0-y_0)\left\{\omega(x_0)-\omega(y_0)\right\}-\omega(T)\right]\Bigr\},\\
H_\pm(\bfp;x_0,y_0) &=&  \frac{1}{2\cosh\omega(T)} \Bigl\{\mp\cosh\left[\omega(x_0)+\omega(y_0)-\omega(T)\right] \nonumber\\
    &&+ \varepsilon(x_0-y_0)\cosh\left[
                    \varepsilon(x_0-y_0)\left\{\omega(x_0)-\omega(y_0)\right\}-\omega(T)
                               \right]\Bigr\}. 
\end{eqnarray}
Here all functions are diagonal in colour space, with colour components being obtained by
specifying the colour components $\phi_i$ and $\phi'_i$ and thus $\omega_i$, with $i=1,\ldots,N$.
Setting $\phi_i=\phi_i'=0$ amounts to the replacement of $\omega_i(t)\rightarrow \omega(\bfp^+)t$ 
and these expressions then reduce to the ones given in Eqs.~(\ref{eq:Gpmcont}) and (\ref{eq:Hpmcont}), 
for the special case with $p_1^+=p_2^+=p_3^+=p^+$.

\subsection{The free quark propagator on the lattice}

A straightforward way to calculate the free quark propagator on the lattice proceeds via the classical 
solution of the lattice Dirac equation, following~\cite{Luscher:1996vw}, where the inhomogeneous boundary conditions
are imposed either at Euclidean times $x_0=0,T$, or appropriately off set by $\pm a/2$.
After spatial Fourier transform, the result can be written in the form
\begin{eqnarray}
S_{\bfp}^\tau(x_0,y_0) &=&\frac{1}{2A(\bfp^+)}\delta_{x_0,y_0} 
+ \gamma_0\tilde{Q}_+H_+^\tau + \gamma_0\tilde{Q}_-H_-^\tau\nonumber\\
&& + \left(-i\tilde{p}^+_k\gamma_k+M(\bfp^+)\right)
\left[\tilde{Q}_+G_+^\tau+ \tilde{Q}_-G_-^\tau\right],
\label{eq:proplat}
\end{eqnarray}
where the coefficient functions are given by
\begin{eqnarray}
 G_\pm^\tau(\bfp;x_0,y_0) &=& \frac{1}{2\tilde{\omega}A(\bfp^+)\cosh\omega T'}
      \Bigl\{-\sinh\omega(|x_0-y_0|-T') \nonumber\\
       &&\hphantom{012345467890123456}\pm \sinh\omega(x_0+y_0-T)\Bigr\},\label{eq:Gpmlatt}\\
   H_\pm^\tau(\bfp;x_0,y_0) &=& \frac{1}{2 A(\bfp^+)\cosh\omega T'}
   \Bigl\{\tilde\varepsilon(x_0-y_0)\cosh\omega(|x_0-y_0|-T') \nonumber\\
 && \hphantom{0123454678901234}\mp \cosh\omega(x_0+y_0-T)\Bigr\}.
\label{eq:Hpmlatt}
\end{eqnarray}
Here, $T'=T+\tau a$ denotes the effective time extent for the fermionic degrees of freedom,
for the orbifold construction with $\tau=0$ and $\tau=\pm 1$ (cf.~Sect.~4).
With 
\begin{equation}
A(\bfp) = 1+ am_0+ \frac12 a^2\hat\bfp^2,
\end{equation}
the one-particle energy $\omega$ is defined by
\begin{equation}
  \sinh\left( \frac{a}2\omega(\bfp)\right) = \left\{\frac{a^2\tilde\bfp^2+\{A(\bfp)-1\}^2}{4A(\bfp)}\right\}^{\frac12}.
\end{equation}
The notation is further specified by
\begin{eqnarray}
  \tilde{\omega}&=&\frac{1}{a}\sinh(a\omega), \\
  aM(\bfp) &=& A(\bfp)-\cosh[a\omega(\bfp)],
\end{eqnarray}
and the above lattice version of the sign function, $\tilde\varepsilon(t)$, is given by
\begin{equation}
   \tilde\varepsilon(t) = \begin{cases} 1 & \text{if $t>0$}, \\
                                    0    & \text{if $t=0$},    \\                              
                                    -1   & \text{if $t<0$}.   \\
                      \end{cases}
\end{equation}
These expressions for the free quark propagator have been checked numerically by
inverting the corresponding Dirac operators at fixed momentum, and semi-analytically by
comparing with Eq.~(\ref{eq:orb2}). For $\tau=0$ a difference with Eq.~(\ref{eq:orb2})
is found for the contact terms at the boundaries, i.e.~for $x_0=y_0=0$ and $x_0=y_0=T$. While the
difference can be easily worked out, the result is not needed here. In fact, these contact terms 
are never referred to provided the fermionic boundary fields are defined as in Sect.~5.